\documentclass{emulateapj}
\usepackage{amsmath}
\usepackage{subfigure}
\usepackage{xcolor}

\usepackage[normalem]{ulem}
\usepackage{lineno}
\begin{document}

\begin{nolinenumbers}
\vspace*{-\headsep}\vspace*{\headheight}
\footnotesize \hfill FERMILAB-PUB-21-303-AE\\
\vspace*{-\headsep}\vspace*{\headheight}
\footnotesize \hfill DES-2020-0632
\end{nolinenumbers}

\title{The Observed Evolution of the Stellar Mass - Halo Mass Relation for Brightest Central Galaxies}
\def\andname{}

\author{
Jesse~B.~Golden-Marx\altaffilmark{1,2},
C.~J.~Miller\altaffilmark{2,3},
Y.~Zhang\altaffilmark{4},
R.~L.~C.~Ogando\altaffilmark{5,6},
A.~Palmese\altaffilmark{4,7},
T.~M.~C.~Abbott\altaffilmark{8},
M.~Aguena\altaffilmark{5},
S.~Allam\altaffilmark{4},
F.~Andrade-Oliveira\altaffilmark{9,5},
J.~Annis\altaffilmark{4},
D.~Bacon\altaffilmark{10},
E.~Bertin\altaffilmark{11,12},
D.~Brooks\altaffilmark{13},
E.~Buckley-Geer\altaffilmark{14,4},
A.~Carnero~Rosell\altaffilmark{15,5,16},
M.~Carrasco~Kind\altaffilmark{17,18},
F.~J.~Castander\altaffilmark{19,20},
M.~Costanzi\altaffilmark{21,22,23},
M.~Crocce\altaffilmark{19,20},
L.~N.~da Costa\altaffilmark{5,6},
M.~E.~S.~Pereira\altaffilmark{3},
J.~De~Vicente\altaffilmark{24},
S.~Desai\altaffilmark{25},
H.~T.~Diehl\altaffilmark{4},
P.~Doel\altaffilmark{13},
A.~Drlica-Wagner\altaffilmark{14,4,7},
S.~Everett\altaffilmark{26},
A.~E.~Evrard\altaffilmark{2,3},
I.~Ferrero\altaffilmark{27},
B.~Flaugher\altaffilmark{4},
P.~Fosalba\altaffilmark{19,20},
J.~Frieman\altaffilmark{4,7},
J.~Garc\'ia-Bellido\altaffilmark{28},
E.~Gaztanaga\altaffilmark{19,20},
D.~W.~Gerdes\altaffilmark{2,3},
D.~Gruen\altaffilmark{29,30,31},
R.~A.~Gruendl\altaffilmark{17,18},
J.~Gschwend\altaffilmark{5,6},
G.~Gutierrez\altaffilmark{4},
W.~G.~Hartley\altaffilmark{32},
S.~R.~Hinton\altaffilmark{33},
D.~L.~Hollowood\altaffilmark{26},
K.~Honscheid\altaffilmark{34,35},
B.~Hoyle\altaffilmark{36},
D.~J.~James\altaffilmark{37},
T.~Jeltema\altaffilmark{26},
A.~G.~Kim\altaffilmark{38},
E.~Krause\altaffilmark{39},
K.~Kuehn\altaffilmark{40,41},
N.~Kuropatkin\altaffilmark{4},
O.~Lahav\altaffilmark{13},
M.~Lima\altaffilmark{42,5},
M.~A.~G.~Maia\altaffilmark{5,6},
J.~L.~Marshall\altaffilmark{43},
P.~Melchior\altaffilmark{44},
F.~Menanteau\altaffilmark{17,18},
R.~Miquel\altaffilmark{45,46},
J.~J.~Mohr\altaffilmark{36,47},
R.~Morgan\altaffilmark{48},
F.~Paz-Chinch\'{o}n\altaffilmark{17,49},
D.~Petravick\altaffilmark{17},
A.~Pieres\altaffilmark{5,6},
A.~A.~Plazas~Malag\'on\altaffilmark{44},
J.~Prat\altaffilmark{14,7},
A.~K.~Romer\altaffilmark{50},
E.~Sanchez\altaffilmark{24},
B.~Santiago\altaffilmark{51,5},
V.~Scarpine\altaffilmark{4},
M.~Schubnell\altaffilmark{3},
S.~Serrano\altaffilmark{19,20},
I.~Sevilla-Noarbe\altaffilmark{24},
M.~Smith\altaffilmark{52},
M.~Soares-Santos\altaffilmark{3},
E.~Suchyta\altaffilmark{53},
G.~Tarle\altaffilmark{3},
and T.~N.~Varga\altaffilmark{47,54}
\\ \vspace{0.2cm} (DES Collaboration) \\
}

\affil{$^{1}$ Department of Astronomy, Shanghai Jiao Tong University, Shanghai 200240, China}
\affil{$^{2}$ Department of Astronomy, University of Michigan, Ann Arbor, MI 48109, USA}
\affil{$^{3}$ Department of Physics, University of Michigan, Ann Arbor, MI 48109, USA}
\affil{$^{4}$ Fermi National Accelerator Laboratory, P. O. Box 500, Batavia, IL 60510, USA}
\affil{$^{5}$ Laborat\'orio Interinstitucional de e-Astronomia - LIneA, Rua Gal. Jos\'e Cristino 77, Rio de Janeiro, RJ - 20921-400, Brazil}
\affil{$^{6}$ Observat\'orio Nacional, Rua Gal. Jos\'e Cristino 77, Rio de Janeiro, RJ - 20921-400, Brazil}
\affil{$^{7}$ Kavli Institute for Cosmological Physics, University of Chicago, Chicago, IL 60637, USA}
\affil{$^{8}$ Cerro Tololo Inter-American Observatory, NSF's National Optical-Infrared Astronomy Research Laboratory, Casilla 603, La Serena, Chile}
\affil{$^{9}$ Instituto de F\'{i}sica Te\'orica, Universidade Estadual Paulista, S\~ao Paulo, Brazil}
\affil{$^{10}$ Institute of Cosmology and Gravitation, University of Portsmouth, Portsmouth, PO1 3FX, UK}
\affil{$^{11}$ CNRS, UMR 7095, Institut d'Astrophysique de Paris, F-75014, Paris, France}
\affil{$^{12}$ Sorbonne Universit\'es, UPMC Univ Paris 06, UMR 7095, Institut d'Astrophysique de Paris, F-75014, Paris, France}
\affil{$^{13}$ Department of Physics \& Astronomy, University College London, Gower Street, London, WC1E 6BT, UK}
\affil{$^{14}$ Department of Astronomy and Astrophysics, University of Chicago, Chicago, IL 60637, USA}
\affil{$^{15}$ Instituto de Astrofisica de Canarias, E-38205 La Laguna, Tenerife, Spain}
\affil{$^{16}$ Universidad de La Laguna, Dpto. Astrof\'isica, E-38206 La Laguna, Tenerife, Spain}
\affil{$^{17}$ Center for Astrophysical Surveys, National Center for Supercomputing Applications, 1205 West Clark St., Urbana, IL 61801, USA}
\affil{$^{18}$ Department of Astronomy, University of Illinois at Urbana-Champaign, 1002 W. Green Street, Urbana, IL 61801, USA}
\affil{$^{19}$ Institut d'Estudis Espacials de Catalunya (IEEC), 08034 Barcelona, Spain}
\affil{$^{20}$ Institute of Space Sciences (ICE, CSIC),  Campus UAB, Carrer de Can Magrans, s/n,  08193 Barcelona, Spain}
\affil{$^{21}$ Astronomy Unit, Department of Physics, University of Trieste, via Tiepolo 11, I-34131 Trieste, Italy}
\affil{$^{22}$ INAF-Osservatorio Astronomico di Trieste, via G. B. Tiepolo 11, I-34143 Trieste, Italy}
\affil{$^{23}$ Institute for Fundamental Physics of the Universe, Via Beirut 2, 34014 Trieste, Italy}
\affil{$^{24}$ Centro de Investigaciones Energ\'eticas, Medioambientales y Tecnol\'ogicas (CIEMAT), Madrid, Spain}
\affil{$^{25}$ Department of Physics, IIT Hyderabad, Kandi, Telangana 502285, India}
\affil{$^{26}$ Santa Cruz Institute for Particle Physics, Santa Cruz, CA 95064, USA}
\affil{$^{27}$ Institute of Theoretical Astrophysics, University of Oslo. P.O. Box 1029 Blindern, NO-0315 Oslo, Norway}
\affil{$^{28}$ Instituto de Fisica Teorica UAM/CSIC, Universidad Autonoma de Madrid, 28049 Madrid, Spain}
\affil{$^{29}$ Department of Physics, Stanford University, 382 Via Pueblo Mall, Stanford, CA 94305, USA}
\affil{$^{30}$ Kavli Institute for Particle Astrophysics \& Cosmology, P. O. Box 2450, Stanford University, Stanford, CA 94305, USA}
\affil{$^{31}$ SLAC National Accelerator Laboratory, Menlo Park, CA 94025, USA}
\affil{$^{32}$ Department of Astronomy, University of Geneva, ch. d'\'Ecogia 16, CH-1290 Versoix, Switzerland}
\affil{$^{33}$ School of Mathematics and Physics, University of Queensland,  Brisbane, QLD 4072, Australia}
\affil{$^{34}$ Center for Cosmology and Astro-Particle Physics, The Ohio State University, Columbus, OH 43210, USA}
\affil{$^{35}$ Department of Physics, The Ohio State University, Columbus, OH 43210, USA}
\affil{$^{36}$ Faculty of Physics, Ludwig-Maximilians-Universit\"at, Scheinerstr. 1, 81679 Munich, Germany}
\affil{$^{37}$ Center for Astrophysics $\vert$ Harvard \& Smithsonian, 60 Garden Street, Cambridge, MA 02138, USA}
\affil{$^{38}$ Lawrence Berkeley National Laboratory, 1 Cyclotron Road, Berkeley, CA 94720, USA}
\affil{$^{39}$ Department of Astronomy/Steward Observatory, University of Arizona, 933 North Cherry Avenue, Tucson, AZ 85721-0065, USA}
\affil{$^{40}$ Australian Astronomical Optics, Macquarie University, North Ryde, NSW 2113, Australia}
\affil{$^{41}$ Lowell Observatory, 1400 Mars Hill Rd, Flagstaff, AZ 86001, USA}
\affil{$^{42}$ Departamento de F\'isica Matem\'atica, Instituto de F\'isica, Universidade de S\~ao Paulo, CP 66318, S\~ao Paulo, SP, 05314-970, Brazil}
\affil{$^{43}$ George P. and Cynthia Woods Mitchell Institute for Fundamental Physics and Astronomy, and Department of Physics and Astronomy, Texas A\&M University, College Station, TX 77843,  USA}
\affil{$^{44}$ Department of Astrophysical Sciences, Princeton University, Peyton Hall, Princeton, NJ 08544, USA}
\affil{$^{45}$ Instituci\'o Catalana de Recerca i Estudis Avan\c{c}ats, E-08010 Barcelona, Spain}
\affil{$^{46}$ Institut de F\'{\i}sica d'Altes Energies (IFAE), The Barcelona Institute of Science and Technology, Campus UAB, 08193 Bellaterra (Barcelona) Spain}
\affil{$^{47}$ Max Planck Institute for Extraterrestrial Physics, Giessenbachstrasse, 85748 Garching, Germany}
\affil{$^{48}$ Physics Department, 2320 Chamberlin Hall, University of Wisconsin-Madison, 1150 University Avenue Madison, WI  53706-1390}
\affil{$^{49}$ Institute of Astronomy, University of Cambridge, Madingley Road, Cambridge CB3 0HA, UK}
\affil{$^{50}$ Department of Physics and Astronomy, Pevensey Building, University of Sussex, Brighton, BN1 9QH, UK}
\affil{$^{51}$ Instituto de F\'\i sica, UFRGS, Caixa Postal 15051, Porto Alegre, RS - 91501-970, Brazil}
\affil{$^{52}$ School of Physics and Astronomy, University of Southampton,  Southampton, SO17 1BJ, UK}
\affil{$^{53}$ Computer Science and Mathematics Division, Oak Ridge National Laboratory, Oak Ridge, TN 37831}
\affil{$^{54}$ Universit\"ats-Sternwarte, Fakult\"at f\"ur Physik, Ludwig-Maximilians Universit\"at M\"unchen, Scheinerstr. 1, 81679 M\"unchen, Germany}

\email{jessegm@sjtu.edu.cn}

\begin{abstract}

We quantify evolution in the cluster scale stellar mass - halo mass (SMHM) relation's parameters using 2323 clusters and brightest central galaxies (BCGs) over the redshift range $0.03 \le z \le 0.60$.  The precision on inferred SMHM parameters is improved by including the magnitude gap ($\rm m_{gap}$) between the BCG and fourth brightest cluster member (M14) as a third parameter in the SMHM relation. At fixed halo mass, accounting for $\rm m_{gap}$, through a stretch parameter, reduces the SMHM relation's intrinsic scatter. To explore this redshift range, we use clusters, BCGs, and cluster members identified using the Sloan Digital Sky Survey C4 and redMaPPer cluster catalogs and the Dark Energy Survey redMaPPer catalog.  Through this joint analysis, we detect no systematic differences in BCG stellar mass, $\rm m_{gap}$, and cluster mass (inferred from richness) between the datsets.  We utilize the Pareto function to quantify each parameter's evolution.  We confirm prior findings of negative evolution in the SMHM relation's slope (3.5$\sigma$) and detect negative evolution in the stretch parameter (4.0$\sigma$) and positive evolution in the offset parameter (5.8$\sigma$). This observed evolution, combined with the absence of BCG growth, when stellar mass is measured within 50kpc, suggests that this evolution results from changes in the cluster's $\rm m_{gap}$.  For this to occur, late-term growth must be in the intra-cluster light surrounding the BCG. We also compare the observed results to Illustris TNG 300-1 cosmological hydrodynamic simulations and find modest qualitative agreement.  However, the simulations lack the evolutionary features detected in the real data.  
\end{abstract}

\section{Introduction}
\label{sec:DESintro}
The stellar mass - halo mass (SMHM) relation is a primary mechanism used to quantify and characterize the galaxy-dark matter halo connection.  Since multiple versions of the SMHM relation exist, we note that for this analysis, we study the brightest central galaxy (BCG) SMHM relation for galaxy clusters ($\rm log_{10}(M_{halo}$ /$(M_{\odot}/h)) \ge 14.0)$, which is the linear correlation that compares the stellar mass of the BCG to the total halo cluster mass, which includes the dark matter.  We do not account for the stellar mass contained within the satellites in this analysis.  The parameters measured as part of the SMHM relation can constrain galaxy formation models, including the amount of AGN feedback in central galaxies \citep{kra14}. The intrinsic scatter in stellar mass at fixed halo mass ($\sigma_{int}$) can constrain processes responsible for quenching star formation in central galaxies \citep{tin17} as well as characterize dark matter halo assembly \citep{gu2016}.  Additionally, the redshift evolution of the slope and scatter provide insight into how BCGs grow and evolve over cosmic time \citep{gu2016,gol19}.
    
BCGs, which solely make up the stellar mass portion of the cluster-scale SMHM relation, are massive, radially extended, elliptical galaxies, that emit a significant fraction of the total light within their host cluster \citep{sch86,jon00,lin04,ber07,lau07,von07,agu11,bro11,pro11,har12}.  BCGs are located near the cluster's X-ray center.  This location, along with their hierarchical formation \citep[e.g.,][]{del07,ose10, van2010}, lead to correlations between their properties and those of their host cluster \citep{jon84,rhe91,lin04,lauer14}.  Additionally, BCGs are surrounded by diffuse halos of intra-cluster light \citep[ICL;][]{zwi33,zwi51}, which are observed to extend radially as far out as $\approx$1Mpc from the center of the BCG \citep{zha18}, and mostly result from the BCG's hierarchical assembly \citep{mur07}. 

BCGs grow ``inside-out'' \citep{van2010}, following a two-phase formation scenario \citep{ose10}; at high redshifts ($z > 2$) the dense core ($r < \approx$10kpc) forms via in-situ star formation, and at lower redshits ($z < 2$), the outer envelope grows hierarchically via major/minor mergers.  The two-phase formation scenario is supported by both observations \citep{van2010,hua18} and dark matter only cosmological simulations that use empirical or semi-analytic models to quantify central galaxy stellar mass growth \citep[e.g.,][]{cro06,del07,guo10,ton12,shankar15}.

As a result of ``inside-out'' growth, all information about the BCG's recent stellar mass growth is contained within the BCG's outer envelope, which extends to the ICL \citep{ose10,van2010}.  Moreover, recent observations suggest that the majority of the BCG's stellar mass may be contained within a radial aperture of 100kpc centered on the BCG \citep{hua18}, and that the 100kpc boundary may represent a transitional regime between the BCG's outer envelope and the ICL \citep{zha18}.  Therefore, when characterizing BCG evolution associated with the parameters of the SMHM relation, it is vital to measure BCG photometry within large radii, as opposed to the more commonly used 20-30kpc aperture radii \citep[e.g.,][]{lin13, zhang16,lin17}
as discussed in \citet{gol19}, referred to as GM$\&$M19.  More specifically, including stellar mass within large radii strengthens the correlation between BCG stellar mass and halo mass \citep{mos18,gol19}.  However, to yield a stronger correlation, one must also incorporate a third parameter related to BCG growth.   

One observational measurement inherently tied to BCG hierarchical growth is m$\rm_{gap}$, the difference in $r-$band magnitude between the BCG and 4th brightest cluster member within half the radius enclosing 200 times the critical density of the Universe ($\rm R_{200}$) \citep{dar10}.  Throughout this paper, we refer to the m$\rm_{gap}$ between the BCG and 4th brightest member as M14.  Using N-body simulations, \citet{solanes16} find that BCG stellar mass linearly increases with the number of progenitor galaxies.  Since the BCG's central location leads to faster merger growth than that of non-central galaxies, as BCGs grow hierarchically, their stellar mass and magnitude increase, while those same parameters for the 4th brightest member galaxy remain the same (unless that galaxy is involved in the BCG merger).  Therefore, BCG growth results in a corresponding increase in the magnitude gap ($\rm m_{gap}$) and yields the correlation between $\rm m_{gap}$ and BCG stellar mass \citep{har12,gol18}.  Thus, it follows that $\rm m_{gap}$ can be thought of a statistical latent parameter within the cluster SMHM relation, as first presented by \citet{gol18}, which from here on is referred to as GM$\&$M18.  Additionally, the correlation between $\rm m_{gap}$ and stellar mass, which results from hierarchical growth, suggests that $\rm m_{gap}$ may be a tracer of formation redshift (GM$\&$M18).  Given that, we use M14 in this analysis, as opposed to alternative $\rm m_{gap}$ measures, because \citet{dar10} find that systems with large M14 measurements, are more efficiently identified as earlier forming systems than those with large values of M12, the $\rm m_{gap}$ between the BCG and 2nd brightest cluster member.

GM$\&$M18 incorporate $\rm m_{gap}$ into the cluster SMHM relation as a linear stretch parameter, which acts to spread the observed range of stellar masses at fixed halo mass. This is just clarifying one of the primary components of the intrinsic scatter in the classic SMHM relation (i.e., without using $\rm m_{gap}$ as a third parameter). In other words, the intrinsic scatter measured using the standard 2-parameter SMHM relation is larger than the intrinsic scatter in the SMHM relation after accounting for the third parameter (e.g., akin to a fundamental plane).  The inferred intrinsic scatter in the SMHM relation found by GM$\&$M18 is less than 0.1dex, which is smaller than previous studies by as much as a factor of two \citep[e.g.,]{gu2016, tin16, zu16,kra14,pil17}. Since the scatter in the SMHM relation is quite small, the other parameters can be more precisely constrained, but only after incorporating the stretch parameter, which measures the strength of the correlation between $\rm m_{gap}$ and stellar mass at fixed halo mass. 

Next, consider the evolution of the SMHM relation over cosmic time.  This evolution can inform us about how BCG's grow over time as well as the fraction of stellar material ejected into the ICL as a result of major/minor mergers. Using empirical models with abundance matching techniques to infer halo masses, \citet{beh13} and \citet{mos13} find that the slope of the SMHM relation increases by a factor of 1.5-2.0 from $z$=1.0 to $z$=0.0, which would suggest that BCGs continue to grow significantly via mergers over this redshift range.  Moreover, \citet{mos13} detect moderate evolution from $z$=0.5 to $z$=0.  In contrast, \citet{pil17} and \citet{eng20}, using the Illustris TNG300-1 cosmological hydrodynamic simulation, measure little change in the slope between $z$=1.0 and $z$=0.0.  In addition to the slope, the expected redshift evolution of the intrinsic scatter, $\sigma_{int}$, has also been investigated in models and simulations \citep{mat17, pil17, gu2016}.  However, as was the case for the slope, there is no consensus between these studies.

Since our analysis accounts for the satellite population via $\rm m_{gap}$, it is also worth highlighting two recent results looking at the evolution within the total stellar mass of DES clusters using the parameter $\mu_{*}$, the sum of the individual galaxy stellar masses weighted by their membership probability.  \citet{pal20} find no evolution in the correlation between $\mu_{*}$ and richness ($\lambda$).  In contrast,  \citet{per20} explicitly accounts for redshift evolution in their relation and finds weak redshift evolution.   However, \citet{per20} note that their evolution is within the accepted uncertainty of the total stellar mass ($\approx$ 0.1 dex).  Therefore, like for the SMHM relation for BCGs, it is currently unclear how the stellar mass of the cluster is evolving. 

Using observational data, prior studies have been unable to constrain the SMHM relation's late time redshift evolution \citep{oli14,goz16,Erfanianfar19}. However, by incorporating $\rm m_{gap}$, GM$\&$M19 placed the first statistically significant observational constraints on the redshift evolution of the slope of the SMHM relation. Over the redshift range $0.03 < z < 0.30$,  the slope of the SMHM relation decreases by $\approx$0.20dex or 40$\%$. To expand upon those results, a primary goal of this paper is to characterize the evolution of the cluster SMHM out to $z \sim 0.6$.

To constrain evolution in the SMHM relation to higher redshifts, we combine the lower redshift Sloan Digital Sky Survey (SDSS) data with  data from the Dark Energy Survey Year 3 (DESY3) release \citep{DES05,DESDATA1,DESDATA2}.  We chose DES data for a few key reasons.  First, tens of thousands of galaxy clusters are identified in DES and the survey is complete out to z$\approx$0.6, significantly deeper than the redshift range probed by SDSS \citep[e.g.,][]{ryk14,alam15}.  While surveys such as Hyper Suprime-Cam Subaru Strategic Program \citep[HSC SSP][]{aih18} or the Atcama Cosmology Telescope \citep[ACT][]{Hil20} may offer a similarly deep (or deeper) redshift coverage, those surveys do not provide a large enough sample of clusters to reduce the statistical uncertainty needed to provide tight constraints on the parameters associated with the SMHM relation using our Bayesian model.  Additionally, the deep DES photometry allows us to accurately measure large aperture photometry for our BCGs.  DES provides a wide field of view around each BCG allowing us to easily determine $\rm m_{gap}$ as well.  Finally, one goal of this analysis is to create a homogeneous data set to study redshift evolution.  Since the redMaPPer algorithm \citep{ryk14,ryk16} has been applied to DES data, and there exists a set of clusters observed by both SDSS and DES, this makes the process of creating a homogeneous sample and determining the associated uncertainty simpler, since the membership restrictions and measurement methods applied to SDSS data can be similarly applied to DES data.

The outline for the remainder of this paper is as follows. In Section~\ref{sec:DES}, we discuss the observational and simulated data (Illustris TNG300-1) used to measure stellar masses, halo masses, and $\rm m_{gap}$ values for our SMHM relation.  In Section~\ref{sec:DESmodel}, we describe the hierarchical Bayesian MCMC model used to evaluate the redshift evolution of the SMHM relation.  In Section \ref{sec:DESresults}, we present our results.  In Section~\ref{sec:DESdiscussion} we discuss our findings and conclude. 

Except for the case of the TNG300-1 simulated data, in which the cosmological parameters are previously defined ($\Omega_{M}$=0.3089, $\Omega_{\Lambda}$=0.6911, H$_{0}$=100~$h$~km/s/Mpc with $h$=0.6774), for our analysis, we assume a flat $\Lambda$CDM universe, with $\Omega_{M}$=0.30, $\Omega_{\Lambda}$=0.70, H$_{0}$=100~$h$~km/s/Mpc with $h$=0.7.

\section{Data}
\label{sec:DES}

To characterize the SMHM relation and its evolution, we require mass measurements of the central galaxies, as well as enough satellite galaxies to infer $\rm m_{gap}$ and the richness of the halo, the latter of which allows for an estimate of the halo's mass. To obtain these measurements over the desired redshift range, we utilize two survey data sets: the Sloan Digital Sky Survey Data Release 12 (SDSS DR12) \citep{alam15} and the Dark Energy Survey Year 3 (DESY3) \citep{DESDATA1,DESDATA2,DESDATA3}, which are briefly summarized below.  

DES is an optical-to-near-infrared photometric survey covering 5,000 deg$^2$ in the South Galactic Cap in the DES $grizY$ bands (for the purpose of this analysis, only the $g-$, $r-$, $i-$, and $z-$bands are used).  In total, over 575 nights of observation were taken over a 6-year period, beginning in 2013.  The observations were taken at the Cerro Tololo Internation Observatory (CTIO) in Chile using the $\approx$3 deg$^2$ CCD Dark Energy Camera \citep[DECam][]{fla15} on the Blanco 4-m telescope.  The data used in this analysis were taken over the first three years of observations.

SDSS is an photometric survey with overlapping spectroscopic data collected by Baryon Oscillation Spectroscopic Survey (BOSS) that covers a footprint in the northern sky of 14,055 deg$^2$ in the SDSS $griz$ bands (for the purpose of this analysis only the $g-$, $r-$, and $i-$bands were used) observed between 1998-2009.  The observations were taken using the Sloan Foundation 2.5-m telescope at Apache Point Observatory in New Mexico.  As in GM$\&$M18 and GM$\&$M19, the data used in this analysis come from SDSS DR12.  The only difference between the data used in this analysis and the prior studies is in the radii within which the SDSS BCG magnitudes are measured, as discussed in Section~\ref{subsec:DESBCGphot}.

The galaxy clusters that are used in this analysis come from the low-redshift SDSS-C4 \citep{mil05}, SDSS-redMaPPer v6.3 \citep{ryk14}, and DES-redMaPPer v6.4.22 $\lambda > $20, volume limited \citep{ryk16} catalogs, where $\lambda$ is the DES richness measurement.  These redMaPPer cluster catalogs have both high purity and high completeness over the redshift and $\lambda$ ranges that we are studying \citep{ryk14,ryk16}.  However, no single cluster catalog individually covers the entire redshift range we aim to study: $0.03 < z < 0.60$.  The SDSS-C4 sample covers $0.03 < z < 0.15$, with $\rm z_{med}=0.08$, the SDSS-redMaPPer sample covers $0.08 < z < 0.30$ and the DES-redMaPPer sample covers $0.20 < z < 0.60$.  Therefore we combine the cluster catalogs to create one parent sample. 

GM\&M19 used clusters in the redshift range $0.08 \le z \le 0.12$ to characterize any differences in the halo masses, richnesses, central galaxy magnitudes, stellar masses, and magnitudes gaps between the SDSS-C4 and SDSS redMaPPer data. By conducting a direct comparison on individual clusters in both data sets, they ruled out systematic differences in the mean observables (e.g., biases) between the two samples which could mimic real evolutionary trends in the SMHM relation. We conduct a similar analysis on an overlapping redshift region for the SDSS-redMaPPer and DES-redMaPPer clusters in this work, described in Section~\ref{subsec:DESuncertainties}.

In our SMHM relation analysis, we constrain the evolution of the parameters with and without redshift binning to emphasize consistency in our statistical analysis.  For the redshift binned analysis, the parent sample of SDSS and DES clusters is divided into 8 redshift bins as shown in Figure~\ref{fig:DES_redshift_dist} and given in the Appendix.  
\begin{figure}[ht]
    \centering
    \includegraphics[width=8cm]{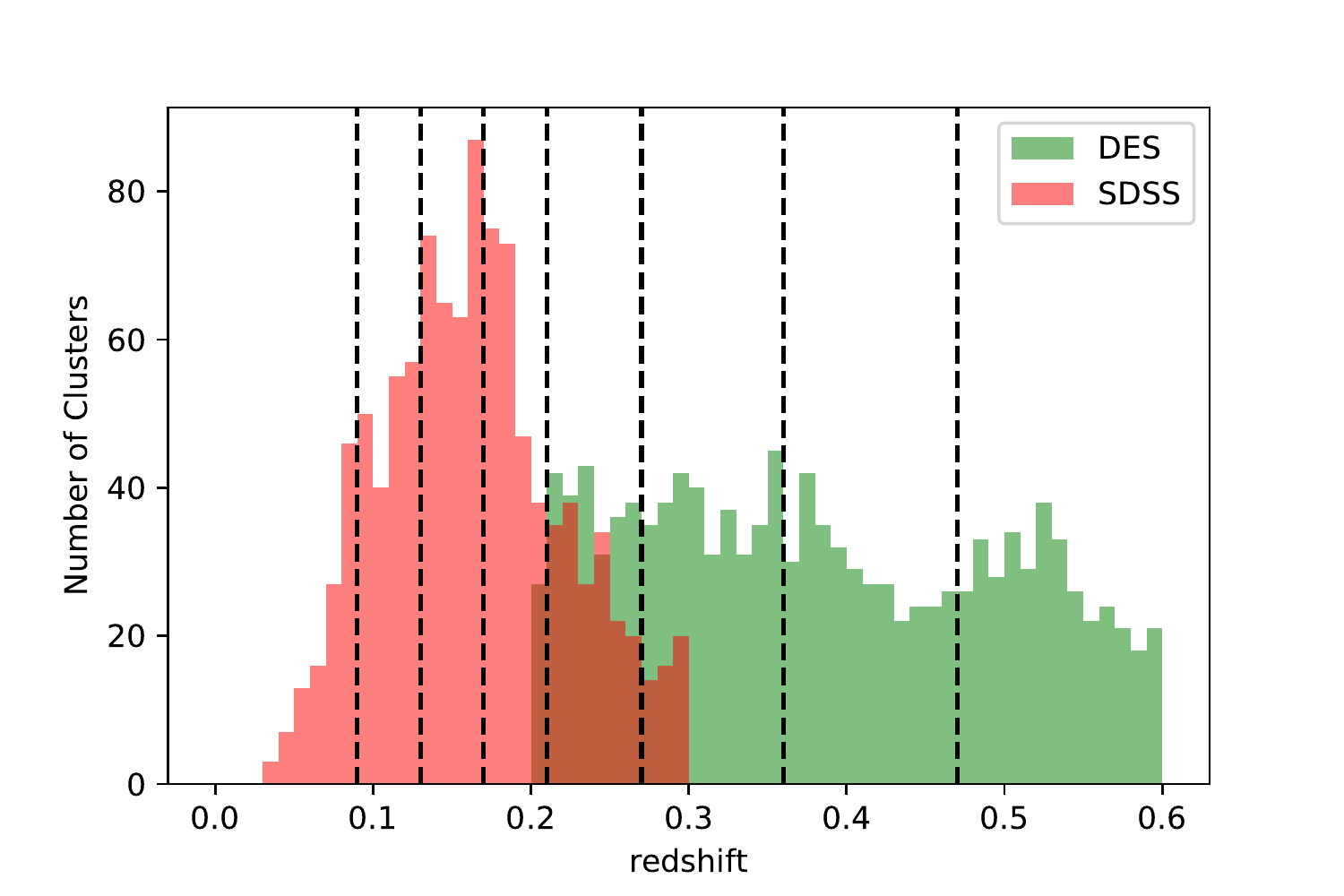}
    \caption{The redshift distribution of the combined sample of SDSS and DES data for this analysis.  The dashed lines represent the edges of the 8 bins used in this analysis.}
    \label{fig:DES_redshift_dist}
\end{figure}

In the following subsections we describe the measurements of our observables: BCG stellar mass, cluster mass (via richness), and $\rm m_{gap}$. We will specifically highlight important differences compared to the measurement approaches used in GM\&M19. In addition to the observables, our Bayesian analysis also requires priors on the measurement uncertainties. We estimate these uncertainties from a bootstrapped Bayesian analysis described in Section~\ref{subsec:DESuncertainties}.

\subsection{BCG identification}
\label{subsec:BCGid}
The BCG in every cluster is identified using a combination of visual identification, magnitudes, color, and redshift. While the BCG is nominally the brightest galaxy at the center of the cluster's gravitational potential well, difficulties in BCG identification arise due to cluster centering accuracy, foreground/background contamination, photometric accuracy, etc.  For the low redshift SDSS-C4 clusters, the BCGs were visually confirmed. For most of the redMaPPer clusters, we use the statistically most probable BCG from the redMaPPer algorithm.  However, in the overlap samples (i.e., clusters that appear in two or more of the three catalogs), we visually confirm the BCG (a thorough discussion of miscentering in redMaPPer is provided in \citet{hos15,zha19}). We note that the BCG must lie within 0.5$\times \rm R_{200,crit}$ of the cluster center (see Section below for details). The photometric algorithms and visual confirmations ensure that the BCGs have similar colors to the rest of the galaxies within the so-called E/S0 ridgeline and that the BCG morphologies exclude disk, disturbed, or merging galaxies.

\subsection{BCG light profiles}
\label{subsec:DESBCGphot}
GM$\&$M19 found that the slope of the SMHM relation is dependent on the radius within which the BCG's stellar mass is measured. The SMHM relation's slope reaches an asymptote when the projected aperture used to estimate the stellar mass within a galaxy is between 60-100kpc. Therefore, to homogeneously infer the slope, we use fixed physical aperture BCG magnitudes as opposed to alternatives such as the Petrosian or Kron magnitude \citep{Petrosian76, Kron80}. 

For the SDSS BCGs, we use the SDSS pipeline processed radial light profiles to measure fixed aperture magnitudes. The DES pipeline does not provide radial aperture photometry. Therefore, we follow the procedure described in \citet{zha18} to measure the DES BCG light profiles.

\begin{figure*}[ht]
    \begin{center}
    \subfigure{\includegraphics[scale=0.5]{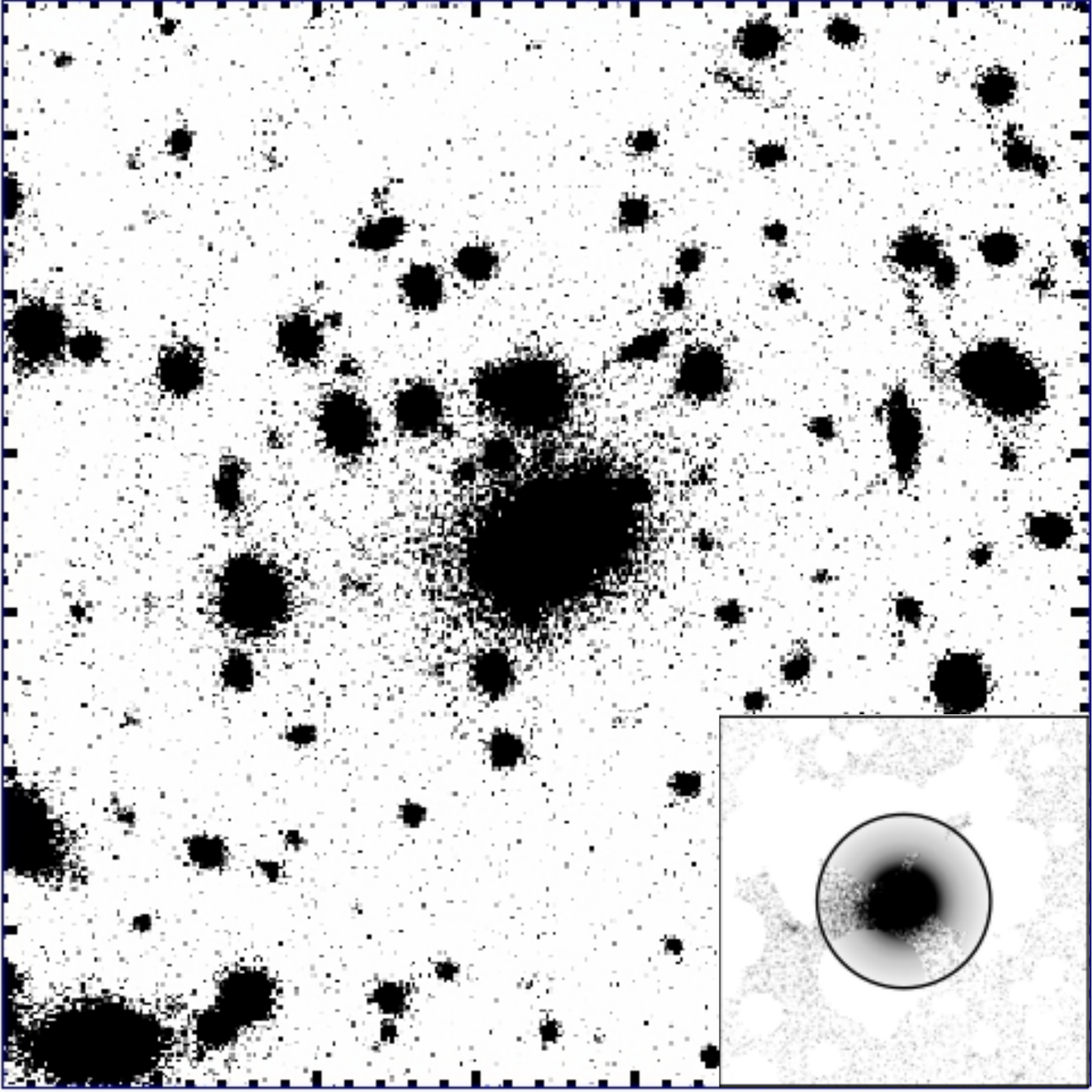}}
    \subfigure{\includegraphics[scale=0.5]{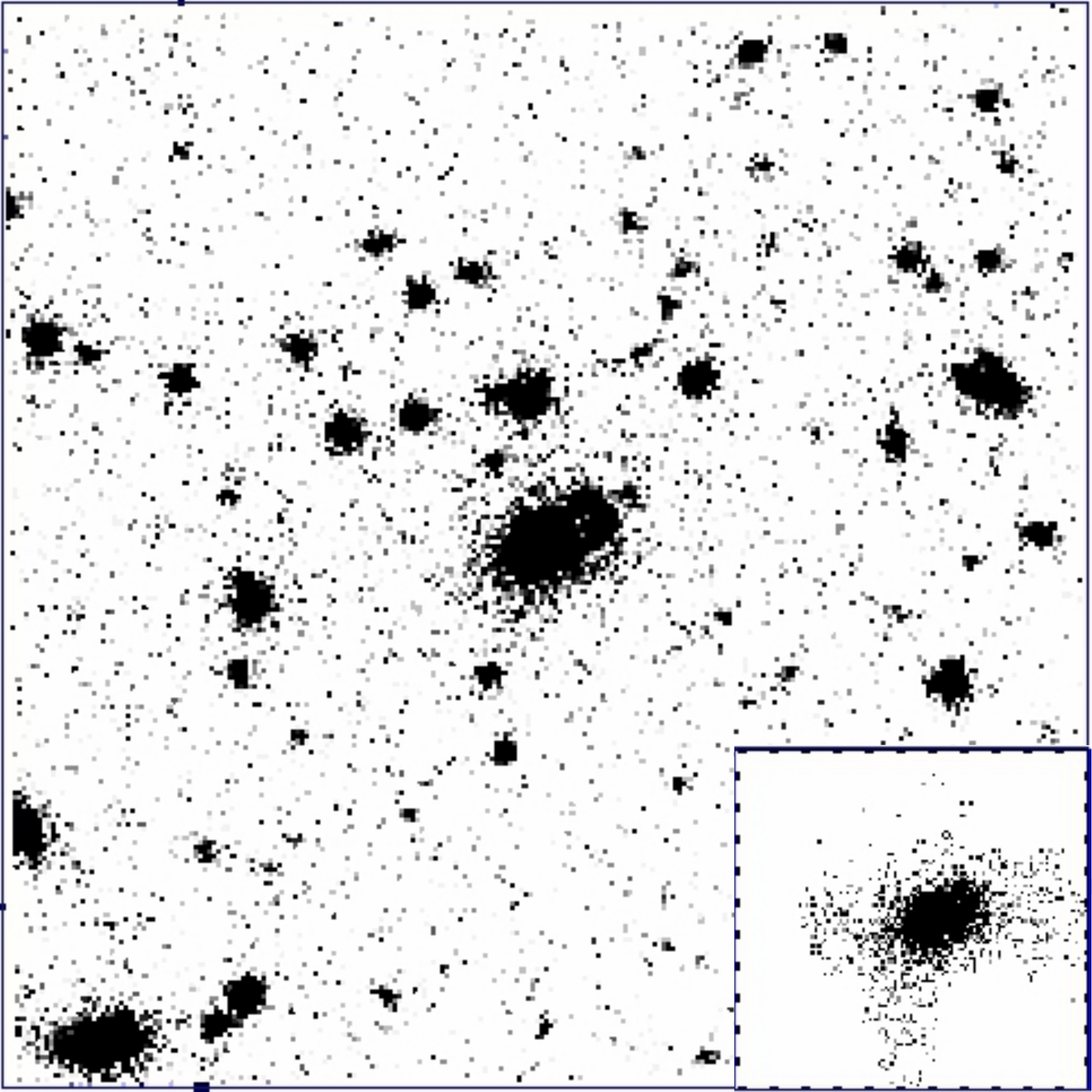}}
    \caption{The left image is a coadded DES cluster, centered on the BCG.   The right image is the SDSS image of the same BCG.  Each image is a 1.5' by 1.5' box centered on the BCG. The inset in both panels is a representation of the BCG after masking. The black  circle is the 50kpc radius, within which we measure the BCG stellar mass. Note that while the DES postage stamp appears more spherical than the SDSS postage stamp, we do not use shape information in this work.}
    \label{fig:DES_coadd}
\end{center}
\end{figure*} 

We coadd and stack the DES individual image frames out to 0.15$^{\circ}$ from the BCG locations.  In Figure~\ref{fig:DES_coadd}, we compare a DES coadded and sky-subtracted image to an SDSS pipeline processed image of the central region for a specific cluster. Unsurprisingly, the DES data reach a much lower surface brightness than the SDSS photometry.  In the SDSS pipeline, the radial light profiles are constructed after nearby objects are masked.  Masking removes the majority of excess light associated with the neighboring galaxies, yielding a clean measurement of the radial light profile centered on the BCG, which includes the ICL. For the DES data, we mask all objects brighter than 30th magnitude in the $i-$band out to a radius of 2.5R$\rm_{kron}$ for each detected object.  In the inset images of Figure \ref{fig:DES_coadd}, we compare the masked and sky subtracted DES BCG to the SDSS Atlas image, the SDSS equivalent. In the DES masked recovered image, the masked pixels are replaced with the radially averaged flux level.

\begin{figure}[ht]
    \centering
    \includegraphics[width=8cm]{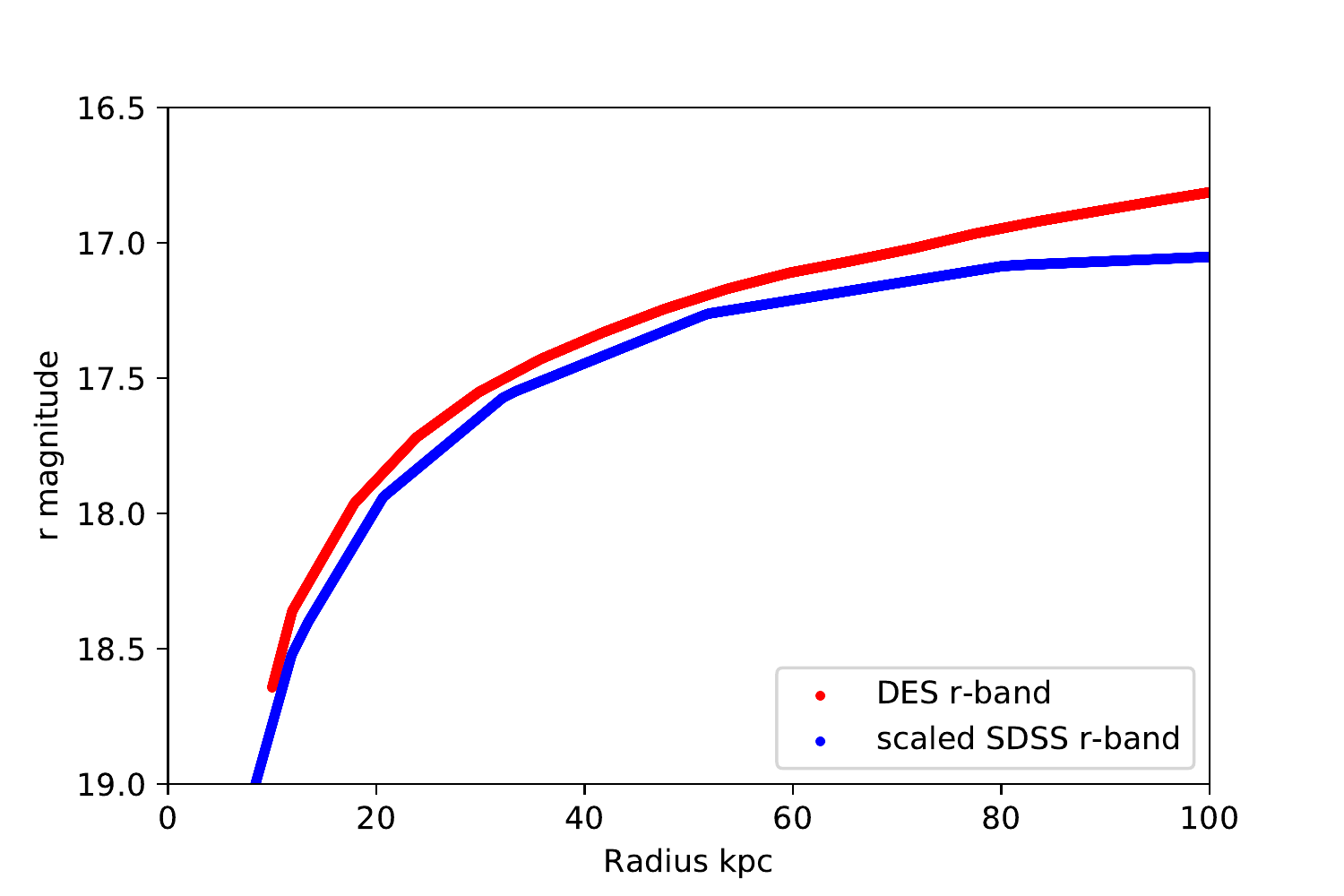}
    \caption{The SDSS and DES $r-$band light profiles for the BCG in Figure \ref{fig:DES_coadd}. For such a comparison, we scaled the SDSS photometry to match that of the DES photometry, since there are slight differences in the wavebands used for the analysis.  The DES profile matches the SDSS within the inner 50kpc region, but then becomes brighter as more light above the background is observed in the deeper DES observations.}
    \label{fig:DES_lightprofile}
\end{figure}

After the mask is applied to the DES co-added image,  we measure the BCG's radial light profile in annuli centered on the BCG.  We subtract a background determined from the median flux at radii beyond 500kpc from the BCG. In Figure~\ref{fig:DES_lightprofile} we compare the $r-$band light profiles for a single BCG in both the SDSS and DES photometry, which shows that while the light measured within the central aperture is very similar, there is more light in the DES photometry compared to SDSS photometry at larger radii.

\begin{figure}[hb]
    \centering
    \includegraphics[width=8cm]{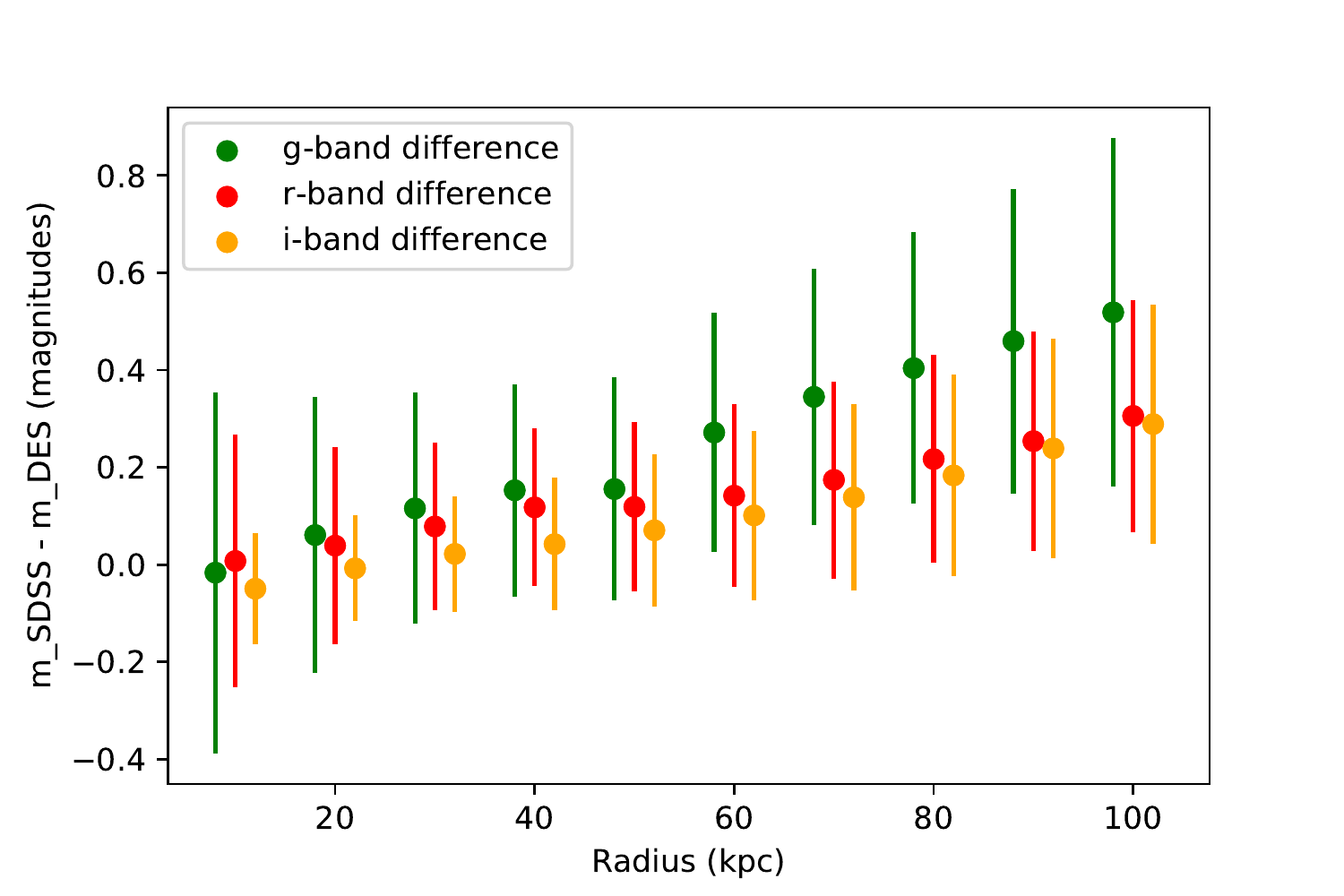}
    \caption{The difference between the SDSS and DES photometric measurements for 48 BCGs identified in both the SDSS and the DES redMaPPer catalogs. We show the cumulative magnitude difference for the $g-$, $r-$, and $i-$bands as a function of the radial aperture. The SDSS and DES photometry begin to diverge at radii $>$ 50kpc, particularly in the $g-$band where differences between the two surveys are larger than the average magnitude measurement error on the BCG magnitude.}
    \label{fig:SDSS_DES_comp}
\end{figure}

We identify 48 BCGs within $0.20 < z < 0.35$ observed in both DES and SDSS that have SDSS light profiles which are above the background to 100kpc. We use this subset of data to quantitatively characterize the radially dependent magnitude differences between DES and SDSS BCG photometry. Beyond 50kpc, the SDSS photometric measurements are consistently fainter than the DES photometry as shown in Figure ~\ref{fig:SDSS_DES_comp}. The differences begin to grow beyond 50kpc, with the DES magnitudes nearly 0.5 magnitudes brighter in the $g-$band compared to SDSS when measured at 100kpc. The differences are less pronounced in the $r-$ and $i-$bands, but large enough to cause concern about using the 100kpc aperture magnitude for BCGs.  Based on this analysis, we choose 50kpc as the BCG aperture magnitudes for the remaining analyses. We use this aperture for all SDSS and DES BCGs.  We note that since we do not use the SDSS $z-$band photometry no comparison is made between the SDSS and DES $z-$band.

As a final test, we compare the colors of the 48 BCGs which exist in both SDSS and DES data. We convert the SDSS magnitudes to DES magnitudes using the available filter curves for each survey \citep[e.g.,][]{alam15,Li16,bur18}.  We then calculate the mean and standard deviation of the difference between the SDSS and DES BCG colors. We find that mean (error) is 0.035 magnitudes with a standard deviation of 0.137 magnitudes. Therefore, we find no bias between the two data sets for the BCG colors.

\subsection{BCG stellar masses}
\label{subsubsec:DESBCGMstar}

We use the observed radial light profiles to measure the projected BCG luminosity/magnitude within a 50kpc aperture. The magnitudes in different bands allow us to estimate the stellar mass.  We follow the same procedure as outlined in GM$\&$M19, summarized here.  For each cluster, redMaPPer assigns every galaxy a membership probability, $\rm P_{mem}$, which is dependent on the cluster's richness, density profile, and background density \citep{ryk14,ryk16}.  The $\rm P_{mem} >0.7$ members are then used to estimate the cluster's photometric redshift, which we use for the BCG. %  \chris{Jesse: what value of P$_{mem}$ is used for the redshift?} 

Given the BCG apparent magnitudes, colors, and redshifts, we then fit a passively evolving spectral model using the SED modelling software package EzGal \citep{man12}. This model fitting allows us to infer a color-based stellar mass.  We assume a \citet{bru03} stellar population synthesis model, a \citet{sal55} Initial Mass Funtion (IMF), a formation redshift of $z$=4.9, and a metallicity of either 0.008, 0.02, or 0.05.  The choice of metallicity for each DES BCG is determined based on which model yielded the lowest chi-squared statistic between the measured and modelled photometry.  We note that GM$\&$M19, found that $> 99\%$ of lower-redshift SDSS clusters are best constrained by the model when a metallicity of 0.008 is used; however, that fraction decreases to 87\% for the high-redshift DES data.  To determine the best fit SED, we use a Bayesian MCMC approach using emcee \citep{for13}, where we treat the absolute magnitude (the EzGal normalization parameter) as a free parameter with a uniform prior. The colors generated by the EzGal model are then compared against the $g-r$ and $r-i$ colors for SDSS and either the $g-r $ and $r-i$ or $r-i$ and $i-z$ colors for DES to determine the absolute magnitude that yields the best fit for our observations.  We use different colors depending on the redshift of the data because model degeneracies can become a problem in the $g-r$ colors at $z > 0.35$.  Additionally, we note that had we chosen either a different IMF or formation redshift, the only impact on our results would be a uniform shift in the stellar mass values, which would only impact the value, and not the evolution in the $\alpha$ parameter.

Based on a comparison between 61 clusters between $0.20 \le z \le 0.35$ in both SDSS and DES (with light profiles out to 50kpc), we find that the mean difference between our stellar mass measurements is 0.04 dex with a standard devation of 0.07. Thus, the difference between the 50kpc SED inferred BCG stellar masses from two independent imaging surveys is statistically consistent with zero.  Therefore, we find excellent agreement between stellar masses estimated for BCGs with both DES and SDSS photometry.  We also test to confirm that $gri$ and $riz$ photometrically determined stellar masses are consistent, except for increased scatter when color degeneracies appear, and find a median difference of $-0.001$ with a standard deviation of $0.013$.

\subsection{DES Cluster Richnesses and Masses}
\label{subsec:DESMhalo}
For the low redshift SDSS-C4 sample, we use the preliminary mass-richness relation from GM$\&$M19, which was shown to have masses that agree with the SDSS-redMaPPer clusters to within 0.1 dex.  For the SDSS-redMaPPer clusters we use the \citet{sim17} mass-richness relation, which is given by Equation~\ref{eq:SDSSm-lambda},
\begin{equation}
    \label{eq:SDSSm-lambda}
    M_{halo}/(h^{-1} M_{\odot}) = 10^{14.344} (\lambda/40)^{1.33}
\end{equation}
and for the DES clusters, we use the mass-richness relation from \citet{mcc19}, which is calibrated for the DES Year 1 redMaPPer data, given by Equation~\ref{eq:DESm-lambda}.
\begin{equation}
    \label{eq:DESm-lambda}
    M_{halo}/(h^{-1} M_{\odot}) = 10^{14.344} (\lambda/40)^{1.356} (\frac{1+z_{red}}{1+0.35})^{-0.30}
\end{equation}
For both equations, M$\rm_{halo}$ is M$\rm_{200m}$, $z\rm_{red}$ is the redMaPPer photometric redshift and $\lambda$ is the redMaPPer defined richness.  We note that both the \citet{sim17} and \citet{mcc19} mass-richness relations have intrinsic scatters associated with the halo mass at fixed richness.  While not shown in Equations~\ref{eq:SDSSm-lambda} or \ref{eq:DESm-lambda}, we account for this scatter in our Bayesian MCMC analysis, as discussed in Sections~\ref{subsec:DESuncertainties} and \ref{sec:DESmodel}.  The primary difference between the two redMaPPer mass-richness relations is the redshift evolution parameter incorporated into the DES redMaPPer \citet{mcc19} version. We note that we are actually using DES Y3 data and that a preliminary analysis of the DES-redMaPPer Y3 richness estimate is consistent when compared to Y1 analysis.  We also compare the redMaPPer richnesses for the 61 clusters which are in both SDSS and DES and find excellent agreement between the two measurements, with a difference of $0.9 \pm 10.1$.  This translates to an offset of $0.01 \pm$ 0.11 dex in units of $\rm log_{10} M_{\odot}$ in halo mass.  

\subsection{ m$\rm_{gap}$}
\label{subsec:DESmgap}
For the low redshift SDSS-C4 clusters, we use all available spectroscopic information to identify the 4 brightest cluster galaxies and measure $\rm m_{gap}$. We use a radius of 0.5$\times$ $\rm R_{200}$, where $\rm R_{200}$ is the radius where the mass density reaches 200$\times$ the critical density. For the SDSS-C4 clusters, $\rm R_{200}$ is determined from the clusters masses (see GM\&M18 for details).

For the redMaPPer clusters in SDSS and DES, spectroscopic completeness is too low to be of any value for membership criteria. Therefore, we use the redMaPPer red-sequence-based galaxy membership criteria to define $\rm m_{gap}$ values. As discussed in GM$\&$M19, we use all galaxies with a redMaPPer $\rm P_{mem} \ge 0.984$ to define the 4th brightest galaxy within half the virial radius. This high $\rm P_{mem}$ value was chosen because it yielded a match between the red sequence color-parameter space density between a sample of clusters identified in both SDSS-C4 and SDSS-redMaPPer. Additionally, our final sample requires all clusters have 4 or more members (including the BCG) within 0.5 $\rm R_{200}$, which we approximate using Equation~\ref{eq:DES_R200} \citep{ryk14,mcc19}:
\begin{equation}
    \label{eq:DES_R200}
    R_{200} \approx 1.5 R_{c}(\lambda)
\end{equation}
where $\lambda$ is the redMaPPer richness, and $\rm R_{c}$ is the redMaPPer cutoff radius, given by Equation~\ref{eq:DES_Rc}:
\begin{equation}
    \label{eq:DES_Rc}
    R_{c}(\lambda) = 1.0 h^{-1}Mpc (\lambda/100)^{0.2}.
\end{equation}
Using this estimate for $\rm R_{200}$, we define M14 as the difference in the $r-$band apparent model magnitude of 4th brightest cluster member with $\rm P_{mem} \ge $0.984 within 0.5$\rm R_{200}$ and the BCG's 50kpc $r-$band apparent magnitude.  We note that this differs from our definition in GM$\&$M19, which used the apparent model magnitude of the BCG.  However, the choice of the 50kpc magnitude ensures consistency between the DES and SDSS BCG magnitudes as previously shown in Figure~\ref{fig:SDSS_DES_comp}, and $\rm m_{gap}$ values.

We use the same sample of 61 clusters as before to compare M14 measurements between the DES and SDSS data and find $\Delta$M14 = $ -0.08 \pm{0.5}$, which is consistent with zero. We note that there is significant scatter in the data when comparing $\rm m_{gap}$ measurements between DES and SDSS, which likely results from different galaxies being identified as the 4th brightest cluster member.  

\subsection{Statistical Uncertainties}
\label{subsec:DESuncertainties}
For each of the observables used in the analysis, we show that there is no statistically significant difference in the measurements between the observables for the different catalogs and survey data. Recall that we use two surveys (SDSS and DES) and three cluster catalogs to make a combined sample which can be analyzed with and without redshift binning over the range $0.03 \le z \le 0.6$.

We will also incorporate estimates for the uncertainties on the BCG stellar mass, cluster (or halo) mass, and M14, in the Bayesian inference of the SMHM relation. Therefore, just as we need to ensure that differences in the measurements do not introduce systematic evolution in the SMHM relation, we also need to ensure that no such biases arise from the uncertainties on the measurements. To address these uncertainties, we take a similar approach as GM\&M19, where we analyze a redshift bin which has data in both the SDSS and the DES surveys.

GM\&M19 used a combined analysis of the SDSS-C4 and SDSS redMaPPer data to ensure that the statistical uncertainties were similar in the redshift bin $0.08 \le z \le 0.12$, where both catalogs have data. To infer the uncertainties on the observables, they conducted a constrained Bayesian analysis by subsampling the  SDSS-redMaPPer clusters to have the same mass distribution as the SDSS-C4 sample (and over the same redshift) and by using strong priors on the four main parameters which describe the richness dependent SMHM relation, the offset $\alpha$, the slope $\beta$, the stretch $\gamma$, and the intrinsic scatter $\sigma_{int}$ in the multivariate linear relation:
\begin{widetext}
\begin{equation}
    log_{10}(M^{BCG}_{*}/(M_{\odot}/h^2)) = \alpha + \beta\times log_{10}(M_{halo}/(M_{\odot}/h)) + \gamma \times M14
\label{eq:SMHMsimple}
\end{equation}
\end{widetext}
Each BCG stellar mass is treated as a draw from a Normal distribution with mean defined by the above equation and an intrinsic scatter (standard deviation) defined as $\sigma_{int}$.  The priors used for $\alpha$, $\beta$, $\gamma$ and $\sigma_{int}$ were also defined as Normal distributions with means and variances from the analysis on the SDSS-C4 clusters (GM\&M18).  Before running the Bayesian analysis, we shift the data by the difference between the minimum and maximum of the stellar mass and halo mass ($x_{pivot} = 14.65$ and $y_{pivot}=11.50$).  Doing this subtraction removes covariance between $\alpha$ and $\beta$. This is a well known and established technique for constraining scaling law parameters \citep[e.g.,][]{roz14,sim17,mcc19}.   We then treated the uncertainties on the SDSS-redMaPPer observables as free parameters and regressed for their values.

There are numerous advantages to using the SDSS-C4 clusters as the initial rung in the redshift ladder. First, the low redshift SDSS-C4 data include cluster masses which can be inferred from both the dynamics (caustic halo masses) and from a mass-richness relation which provides a self-consistent estimate of the mass uncertainties on the richness based masses (see GM\&M19 for more details). Second, the high spectroscopic completeness of the low redshift clusters minimizes (or eliminates) foreground/background contamination in the $\rm m_{gap}$ measurement. Third, there exist multiple simulation-based mock galaxy samples which mimic the SDSS main galaxy sample. These mock galaxies allow for alternative estimates on the uncertainties of the BCG stellar masses, membership, $\rm m_{gap}$, and cluster masses.  Moreover, by using the results from the SDSS-C4 as the initial rung, we ensure consistency between our measured values, in particular $\sigma_{int}$ and those of prior studies.  In GM$\&$M18, we found excellent agreement between our $\sigma_{int}$ when $\rm m_{gap}$ was not accounted for and other prior results \citep[e.g.,][]{tin16,pil17,kra14}.  This consistency was further reproduced in GM$\&$M19 (see Table 2), which leads to our choice here. 

We follow the same procedure described above to calibrate the uncertainties for the DES BCG stellar masses and $\rm m_{gap}$ values. The SDSS-redMaPPer and DES-redMaPPer samples overlap within $\sim$hundred square degrees and in the redshift range $0.206 < z < 0.30$. We then create a subsample of DES-redMaPPer clusters that matches the redshift range of the SDSS-redMaPPer clusters. We label these two cluster samples as SDSS-Calibration and DES-Calibration in the Appendix.

We note that we have remeasured the SDSS-redMaPPer BCG stellar masses using a 50kpc aperture. Therefore, we first conduct a Bayesian regression analysis on the SDSS-Calibration data to infer the SMHM parameters in equation \ref{eq:SMHMsimple}. We use the same uncertainties as given in GM\&M19 and follow the same algorithm described there. However, there are differences in the data, which is why we conduct this analysis again. Besides the smaller radius used to estimate stellar mass, the SDSS-redMaPPer BCG sample is larger since more BCGs have light profiles measured out to 50kpc than to 100kpc. We find that the fitted parameters of the SDSS-Calibration sample are nearly identical to those from GM\&M19 and provide the results of this analysis in the Appendix. 

We expect differences in the uncertainties between the SDSS and DES data for two reasons. First, the DES data are of much higher quality and depth, leading to a higher signal-to-noise at a fixed aperture in the light profiles (on average). Second, the deeper DES data should make it more difficult to identify the 4th brightest galaxy (on average), often located in the central region of these higher redshift clusters. The former should lead to a reduced uncertainty on the stellar masses (relative to SDSS) and the latter should lead to a larger scatter in the $\rm m_{gap}$ measurements.

We conduct the simplified (no evolution) Bayesian analysis on the DES-Calibration sample and use the SDSS-Calibration posterior distributions for $\alpha$, $\beta$, $\gamma$, and $\sigma_{int}$ as strong priors. We regress for the mean errors associated with $\rm m_{gap}$ and the BCG stellar masses in the DES data.  We find that the DES stellar mass uncertainty is best fit to a value of 0.06 dex, which is smaller than the SDSS-redMaPPer BCGs (0.08 dex), consistent with expectations. We also find that the uncertainties associated with $\rm m_{gap}$ have gone up compared to the SDSS clusters to 0.31 (from 0.15 magnitudes). This is likely due to the DES photometry identifying more objects in the cluster core (see Figure \ref{fig:DES_coadd}) than SDSS-redMaPPer as well as the lower spectroscopic completeness of the DES training set.  We treat the halo mass errors as a fixed value identical to what was found for the SDSS-BCGs in GM\&M19 (0.087 dex) because the halo masses are determined from identical mass-richness relations and since we find no differences in the richness measures for the overlap sample (see Section \ref{subsec:DESMhalo}).  We note that as discussed in GM$\&$M19, our chosen value, 0.087 dex was determined as the result of a joint analysis where the parameters for the SMHM relation were simultaneously determined for a sample with halo masses estimated by both richness and the caustic phase-space technique.  This value in scatter in halo mass at fixed richness corresponds to 0.20 when a natural log scale is used instead of a log$_{10}$ scale.  Therefore, our measured uncertainty has excellent agreement with the results presented in \citet{rozo2015}, which finds the value to be between 0.17-0.21.

We do one final test to ensure that our Bayesian-inferred uncertainties on $\rm m_{gap}$ and the BCG stellar masses in the DES-redMaPPer sample are sensible.  We fix those uncertainties to the values inferred by the prior analysis and we let the SMHM relation parameters $\beta$, $\gamma$, and $\sigma_{int}$ be free in the Bayesian regression. We note that we fix $\alpha$ to the value measured from the SDSS-Calibration, our remeasurement of the results from GM$\&$M19. This is because of the strong degeneracy between $\alpha$ and $\gamma$. The results are presented in rows 2 and 3 of the Appendix. We also report the results from GM\&M19 in row 1 of Table~\ref{tab:DESsim_comp}.  We note that the original results from GM\&M19 use a slightly different pivot point for the data.  However, despite this difference, we find 1$\sigma$ agreement between the GM$\&$M19 measurements and our SDSS and DES-Calbration measurements for all parameters except $\alpha$, which is impacted most significantly by the offset and is a 1.5$\sigma$ difference.

In summary, our goal in this subsection has been to calibrate the uncertainties on the observables in a fixed redshift bin. By ensuring this and also that the mean values of the observables are identical where we expect them to be (i.e., Sections \ref{subsec:DESBCGphot}, \ref{subsubsec:DESBCGMstar},  \ref{subsec:DESMhalo} and \ref{subsec:DESmgap}) we avoid the possibility of inferring evolution where there is none. After quantifying those uncertainties, we test the model to ensure that our conclusions remain unchanged from GM\&M19, which is the case. Next, we combine the SDSS and DES data into a single catalog to search for evolution in the SMHM parameters.

\subsection{DES Final Sample}
\label{subsec:DESFinal}

We analyze how the SMHM relation evolves with redshift using two approaches.  First, we divide our data, incorporating the SDSS-C4, SDSS-redMaPPer, and DES-redMaPPer data, into 8 bins sorted by redshift and measure our Bayesian MCMC posteriors for each bin with redshift evolution parameters set to 0.0.  The redshift range of those bins is given in the Appendix and shown visually in Figure~\ref{fig:DES_redshift_dist}.  Second, we incorporate redshift evolution using four additional parameters described in Section~\ref{sec:DESmodel} and fit over SDSS and DES clusters.  We note that we include all clusters, considered in GM$\&$M19, not just those in the final sample (i.e., we include those clusters that were previously removed as a result of our completeness analyses, since the low-redshift analysis is redone here).  In total, this homogeneous multi-survey data set covers the redshift range $0.03 \le z \le 0.60$.  We note that there is spatial overlap between each of these surveys.  To account for this, we remove galaxies that appear in multiple surveys.  We keep data from DES over SDSS since the photometry is deeper and we keep the SDSS-C4 over SDSS-redMaPPer because of the more stringent red sequence cluster membership.

Following this initial selection criterion, our sample consists of 1172 SDSS clusters, and 1564 DES clusters with halo masses greater than $10^{14} \rm M_{\odot}/h$, which yields a total sample of 2736 clusters.  This number does not account for further halo mass limits; however, as in GM$\&$M19, we expect this total sample of data to have differing halo mass lower limits as we move to higher redshift ranges.  We also check for m$\rm_{gap}$ incompleteness since both SDSS and DES are flux-limited surveys.

For each bin, as in GM$\&$M18 and GM$\&$M19, we apply a $\rm m_{gap}$ completeness criterion based on the binning of the BCG and 4th brightest galaxy's absolute magnitudes against the BCG's apparent magnitude and $\rm m_{gap}$ to determine the apparent magnitude limit of the sample (a redshift dependent limit) \citep{col89, gar99, lab10, tre16, gol18}.  Additionally, since the halo mass distribution can be approximated as Gaussian, the peak indicates the mass that the sample becomes incomplete.  However, we apply a lower halo mass cut located at the halo mass where the amplitude of the binned halo mass distribution decreases to 70$\%$ of the peak value to ensure high completeness out to higher redshifts.  This halo mass criterion is conservative and results in a redMaPPer richness threshold of $\approx 22$, well above the detection limit.  However, when combined with the $\rm m_{gap}$ completeness analysis, these cuts reduce our available sample down to 2323 clusters, a reduction of $\sim15.1\%$.  Slightly more restrictive (higher) halo mass lower limits do not impact our final results.  Of those 2323 clusters in our final sample, 1062 come from SDSS and 1261 come from DES.

\begin{figure}
    \centering
    \includegraphics[width=8cm]{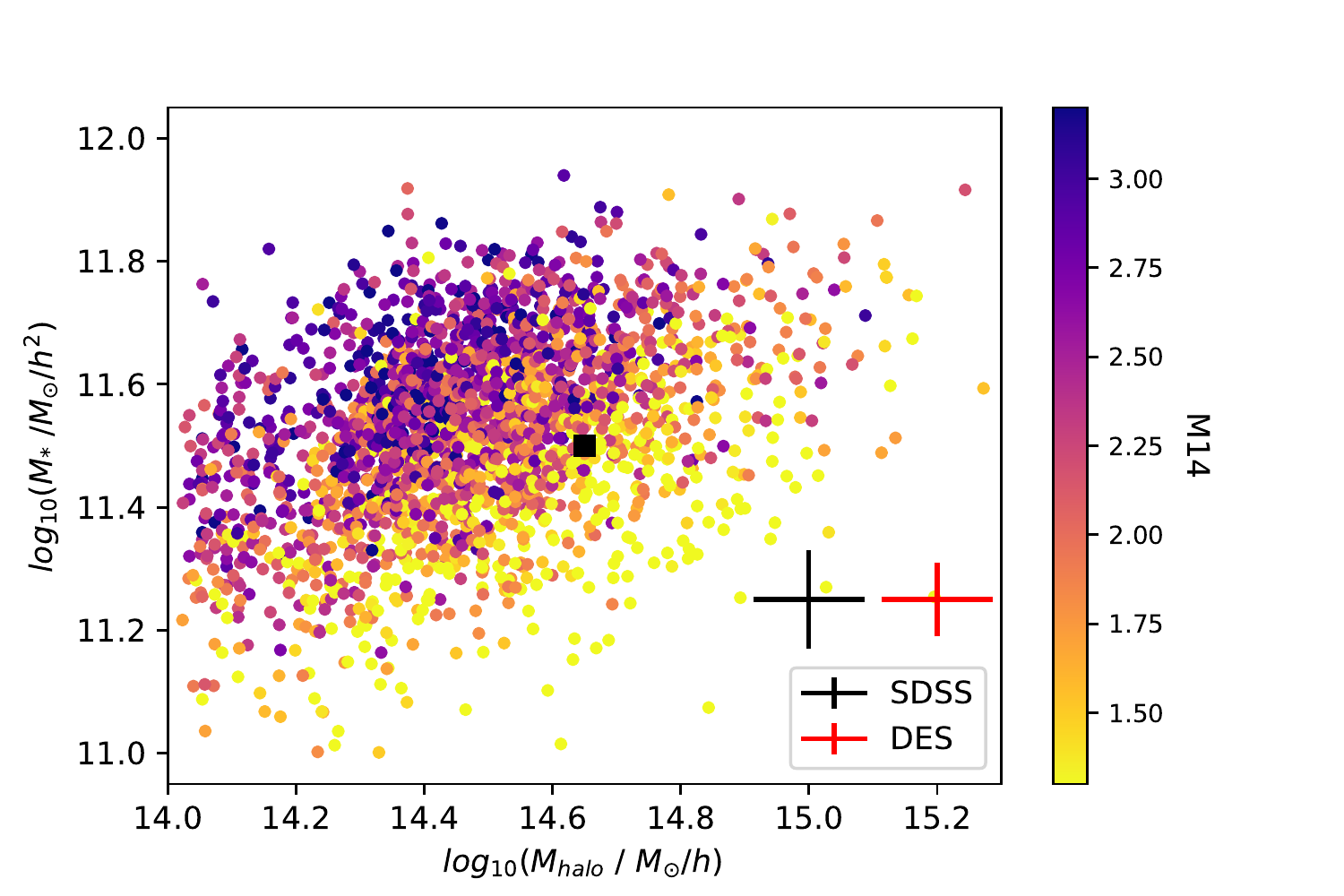}
    \caption{SMHM-M14 Relation for the SDSS-C4, SDSS-redMaPPer, and DES-redMaPPer Samples colored via M14. We see that a stellar mass - $\rm m_{gap}$ stratification continues to persist when measured out to high redshifts.  The black cross represents the error in halo mass, 0.087 dex, and stellar mass, 0.08 dex for the SDSS data and the red cross represents the error in halo mass, 0.087 dex, and stellar mass, 0.06 dex for the DES data.  The black square represents the pivot point that is used in our Bayesian analysis.}
    \label{fig:DES_SMHM_redshift}
\end{figure}

Figure~\ref{fig:DES_SMHM_redshift} visualizes our final sample and shows the 50kpc stellar masses versus the halo masses, color coded by M14. We also show the error bars on each set of survey data (see Section \ref{subsec:DESuncertainties}).  Figure~\ref{fig:DES_SMHM_redshift} includes the entire final sample (following all halo mass and $\rm m_{gap}$ completeness cuts) as described,
and spans the redshift range $0.03 < z < 0.60$, as show in Figure~\ref{fig:DES_redshift_dist}.  Since the dependence on M14 is evident, it is clear that the previously detected stellar mass - M14 stratification continues to exist at higher redshifts.

\subsection{Simulated Data}
\label{subsec:DESsims}
For the simulated analysis of this study, we examine the evolution of the SMHM - $\rm m_{gap}$ relation using the magneto-hydrodynamical cosmological galaxy formation simulations from the IllustrisTNG\footnote{http://www.tng-project.org/} suite \citep{Nel18,pil18,spr18}.  Specifically, we use the Illustris TNG300-1 simulation \citep{pil17} with snapshots analyzed at redshifts of 0.08, 0.11, 0.15, 0.18, 0.24, 0.31, 0.40, and 0.52 (snapshots 92, 90, 87, 85, 81, 77, 72, and 66) the redshifts which best match the median of our 8 binned samples given in the Appendix.  For this comparison, we identify the 260 clusters with $\rm log10(M_{200m}/(M_{\odot}/h)) > 14.0$ in the redshift 0.0 snapshot and use these same clusters in the higher redshift bins.  

For each simulation box, we use 3D information provided directly from each snapshot of the TNG300-1 simulation, including the M$\rm _{200m}$ halo masses, measured within $\rm R_{200} \times \rho_{m}$; the galaxy positions, x, y, z; $\rm R_{200}$; and the magnitudes.  Cluster membership is determined using positional information (x, y, z) and a fit to the red sequence, such that cluster member candidates within 0.5~$\rm R_{200}$, and within 2$\sigma$ from the red sequence, are identified as members.  

\begin{figure}
    \centering
    \includegraphics[width=8cm]{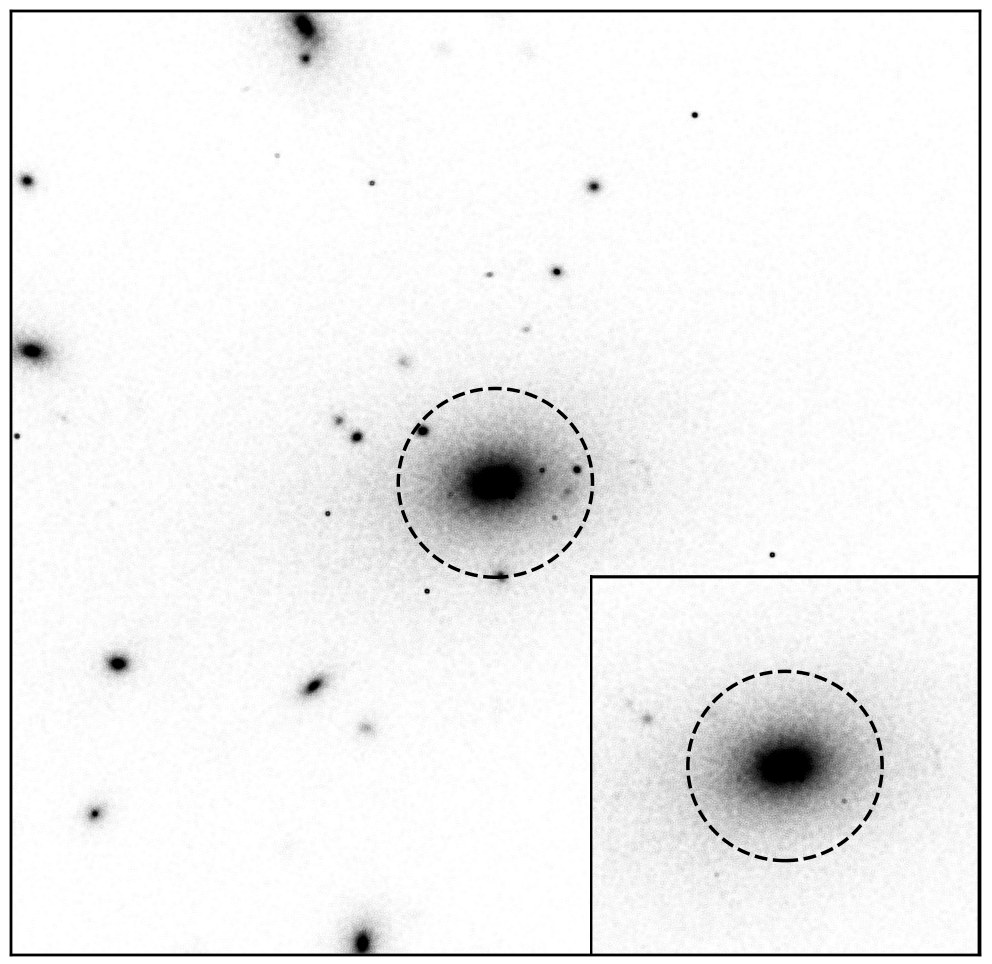}
    \caption{An example of a TNG300-1 image showing the stellar particle information of the halo, which shows that there are many satellite galaxies near to the BCG.  The primary image is the inner 500kpc centered on the BCG.  The insert shows the stellar particle information for just the inner 100kpc of the subhalo containing the BCG.  The circle represents the 50kpc aperture.}
    \label{fig:TNG_image}
\end{figure}

For the observed data, we use a 50kpc aperture on the deblended BCG (see Section \ref{subsec:DESBCGphot} and \ref{subsubsec:DESBCGMstar}). In the simulation, the parent halo and sub-halo have been identified using Subfind \citep{Springel2001}. We therefore use a fixed 50kpc physical aperture for the BCG subhalo alone to calculate the BCG stellar masses.  In other words, we use the Subfind deblending algorithm to separate the stellar components of the BCG from the halo satellite galaxies. An example of one such projected image of a TNG 300-1 BCG is shown in Figure~\ref{fig:TNG_image}.  

\begin{figure}
    \centering
    \includegraphics[width=8cm]{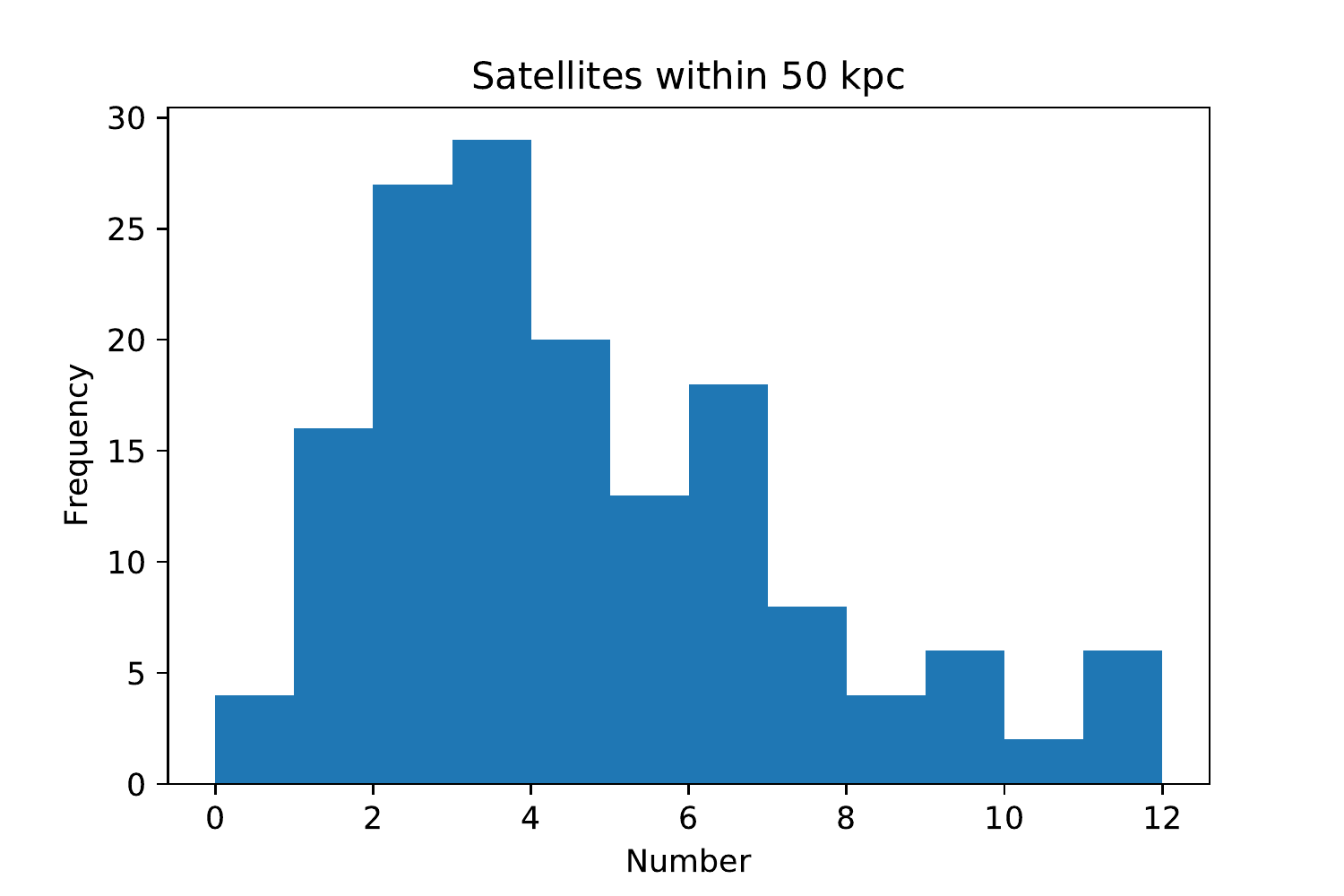}
    \caption{The distribution of the number of satellite galaxies within 50kpc of the TNG300-1 BCGs, which highlights that either masking or using the particle information directly is necessary to estimate stellar masses of TNG 300-1 BCGs. }
    \label{fig:TNG_withinradius}
\end{figure}
The use of a simulation allows us to quantify the impact of deblending in a controlled fashion, since the 3 dimensional positions of the satellites are known. In Figure~\ref{fig:TNG_withinradius}, we show that all but 4 BCGs have a close companion within the 50kpc aperture. The peak of the distribution is around 3 or 4 satellites per BCG, with some BCGs being in very crowded but very localized environments. Proper deblending is therefore necessary to study the BCG itself.  However, we note that as discussed in \cite{zha18}, these satellites are not included in the ICL that is present and which is included in our measurements of the BCG stellar masses in both the data and the simulations. However, we note that stellar particles that may be observationally identified as part of the ICL, but in simulations are associated with the satellite galaxies are thus not incorporated into our stellar mass measurement.  

We use the snapshot information for each BCG subhalo in each of the 8 redshift snapshots.  For M14, we follow a similar procedure to what is done in GM$\&$M19, where M14 is the difference between the $r-$band model magnitude of the 4th brightest member and the $r-$band model magnitude of the TNG300-1 BCG. We do not apply a completeness criterion to our simulated data.  However, to make our approaches between the data and the simulation as homogeneous as possible we apply a halo mass completeness limit, in the same manner as described in Section~\ref{subsec:DESFinal}, which accounts for the fact that the 260 most massive clusters at $z$=0.0 are not guaranteed to be the most massive at each of the higher redshift snapshots because each dark matter halo follows a unique accretion history. 

\section{Bayesian MCMC model}
\label{sec:DESmodel}

We use a similar hierarchical Bayesian MCMC approach to what is described in GM$\&$M19 to determine the values of $\alpha$, $\beta$, $\gamma$, and $\sigma_{int}$ given in Equation~\ref{eq:SMHMsimple}.   Any changes in the underlying model are designed to improve the efficiency of our analysis.  The Bayesian formalism works by convolving prior information for a selected model with the likelihood of the observations given that model, to yield the probability of observing the data for the model, the posterior distribution up to a normalization constant called the Bayesian evidence. 

To determine the posterior distributions for each parameter in the SMHM relation, our MCMC model generates values for the observed stellar mass, halo mass, and $\rm m_{gap}$ values at each step in our likelihood analysis, which are directly compared to our observed measurements.

\subsection{The Observed Quantities}
\label{subsubsec:DESnew_model_observed}

We model the log$\rm_{10}$ BCG stellar masses ($y$), log$\rm_{10}$ halo masses ($x$), and M14 values ($z$) as Normal distributions with mean values (locations) taken from our observed/measured results. The standard deviations associated with each point are taken from  the uncertainties on each measurement, which are determined using the sample of clusters observed by SDSS and DES and include an estimate of the observational uncertainty ($\sigma_{x_{0}}$, $\sigma_{y_{0}}$, $\sigma_{z_{0}}$) as well as a stochastic component from a beta function, $\beta(0.5,100)$ (GM$\&$M18), which allows for additional uncertainty on the observational errors.  These errors are treated statistically in the Bayesian model as free nuisance parameters $\sigma_{x}$, $\sigma_{y}$, and $\sigma_{z}$.

\subsection{The Unobserved Quantities}
\label{subsec:DESnew_model_unobserved}

Our aim is to constrain the parameters of the SMHM relation: the offset, slope, stretch, and intrinsic scatter ($\alpha$, $\beta$, $\gamma$, and $\sigma_{int}$) as a function of redshift.  In GM$\&$M19, we modeled the evolution of these parameters as power-laws $(1+z)^{n_i}$, where $n_i$ defined the amount and shape of the redshift evolution for each of the four model parameters, $\alpha, \beta, \gamma, \sigma_{int}$. 

In this work, we extend the data from the limit of $z \sim 0.3$ to $z \sim 0.6$ using the deeper DES data. We initially tried the same power-law parameterization as in GM$\&$M19. However, we found that a simple power-law could not accurately capture the flattening of the slope at $z > 0.3$ to a single value, while simultaneously having the sharp increase in the slope at low redshift. We explored numerous functional forms and found that the Pareto function, given in Equation~\ref{eq:Pareto}, has the correct shape over the range of data explored in this work.
\begin{equation}
f(a)=\left\{ \begin{array}{ll} 
    \rm{Constant} + n_{i}(a_m^{n_i} / a^{n_i+ 1}) & \hspace{5mm} a \ge a_m \\
    f(a_m) & \hspace{5mm} a < a_m \\
    \end{array} \right.
\label{eq:Pareto}
\end{equation}
This is a Pareto Type I distribution, which is characterized by a scale parameter $a_m$ and a shape parameter $n_i$, which in our case defines the strength of the evolution for $\alpha$, $\beta$, $\gamma$ and $\sigma_{int}$.

A second change from GM$\&$M19 is that we evolve against lookback time as opposed to each cluster's photometric redshift, given that from an astrophysical perspective, stellar evolution is characterized by time as opposed to the universe's scale factor.  We choose $a_m = 0.4$Gyr, which is a redshift of $\sim 0.03$, the lowest redshift our data probe. Below this lower limit, the Pareto distribution is a constant fixed to the value at its lowest data point. The Pareto distribution also asymptotes to a constant at large $a$ (high redshift or lookback time), which we treat as a free nuisance parameter in the analysis. In other words, this is the constant in Equation \ref{eq:Pareto}.
\begin{deluxetable*}{ccc}
	\tablecaption{Bayesian Analysis Parameters for the Combined SDSS-C4, SDSS-redMaPPer, and DES-redMaPPer Sample}
	\tablecolumns{3}
	\tablewidth{0pt}
	\tablehead{\colhead{Symbol} & 
	\colhead{Description} & 
	\colhead{Prior}
	} 
\startdata
$\alpha_{0}$ & The offset of the SMHM relation & $\mathcal{U}$(-20,20) \\ 
$\beta_{0}$ & The high-mass power law slope & Linear Regression Prior \\ 
$\gamma_{0}$ & The stretch parameter, which describes the stellar mass - M14 stratification & Linear Regression Prior \\ 
$\sigma_{int0}$ & The uncertainty in the intrinsic stellar mass at fixed halo mass & $\mathcal{U}(0.0,0.5)$\\ 
$y_{i}$ & The underlying distribution in stellar mass & Equation~\ref{eq:DESSMHM_redshift} \\ 
$x_{i}$ & The underlying halo mass distribution & $\mathcal{N}$(14.23,$0.18^2$)\\ 
$z_{i}$ & The underlying $\rm m_{gap}$ distribution & $\mathcal{N}$(2.51,$0.62^2$)\\ 
$n_{\alpha}$ & The shape parameter associated with the evolution of $\alpha$ & $\mathcal{U}(-5.0,5.0)$\\ 
$n_{\beta}$ & The shape parameter associated with the evolution of $\beta$ & $\mathcal{U}(-5.0,5.0)$\\
$n_{\gamma}$ & The shape parameter associated with the evolution of $\gamma$ & $\mathcal{U}(-5.0,5.0)$\\ 
$n_{\sigma}$ & The power law associated with the evolution of $\sigma_{int}$ & $\mathcal{U}(-20.0,20.0)$\\ 
$\sigma_{y_{0i}}$ & The uncertainty between the observed stellar mass and intrinsic stellar mass distribution & 0.08 or 0.06 dex\\ 
$\sigma_{x_{0i}}$ & The uncertainty associated with the mass-richness relation & 0.087 dex \\ 
$\sigma_{z_{0i}}$ & The uncertainty between the underlying and observed m$_{gap}$ distribution & 0.15 or 0.31 \\
  & & \\ 
\caption{$\mathcal{U}(a,b)$ refers to a uniform distribution where a and b are the upper and lower limits.  The linear regression prior is of the form $-1.5 \times log(1+value^2)$.  $\mathcal{N}(a,b)$ refers to a Normal distribution with mean and variance of a and b.  Additionally, we note that for $x_{i}$ and $z_{i}$, the means and widths given in this table are example values belonging to the lowest redshift bin.} 
\enddata
\label{tab:DESbayes}
\end{deluxetable*}

Using the Pareto distribution, we model the cluster portion of the SMHM relation linearly as: 
\begin{widetext}
\begin{equation}
\label{eq:DESSMHM_redshift}
    y_{i}=
     \underbrace{(\alpha_0+((n_\alpha)*(0.4^{n_\alpha})/(t)^{1+n_\alpha})}_\text{\large evolving offset, $\alpha(t)$} + \underbrace{(\beta_0+((n_\beta)*(0.4^{n_\beta})/(t)^{1+n_\beta})}_\text{\large evolving slope, $\beta(t)$}*x_{i}+
     \underbrace{(\gamma_0+((n_\gamma)*(0.4^{n_\gamma})/(t)^{1+n_\gamma})}_\text{\large evolving stretch, $\gamma(t)$}*z_{i},
\end{equation} 
\end{widetext}
where $x, y, z$ are the observed halo masses, BCG stellar masses, and $\rm{m}_{\rm{gap}}$ values and $t$ is the lookback time, calculated using the photometric redshift.  $\alpha_0$, $\beta_0$, $\gamma_0$ are free parameter offsets which are asymptotic at high lookback time (high redshift). Note that for zero evolution, equation \ref{eq:DESSMHM_redshift} reverts to equation \ref{eq:SMHMsimple}.  We also assume a Gaussian likelihood form, with $\sigma_{int}(t)$ that evolves with redshift: $\sigma_{int_0}+((n_\sigma)*(0.4^{n_\sigma})/(t)^{1+n_\sigma})$.  $n_{\alpha}$, $n_{\beta}$, $n_{\gamma}$, and $n_{\sigma}$ measure the redshift evolution of $\alpha$, $\beta$, $\gamma$, and $\sigma_{int}$ respectively.  

This model is nested.  Thus, for the redshift binned samples, these $n$ parameters are set to 0.0, and as in GM$\&$M19, the zero redshift evolution model from GM$\&$M18 is returned.  This approach allows us to interpret how much better a given model is (e.g., with redshift evolution vs. without) using only the posterior distribution.  We also note that in Equation~\ref{eq:DESSMHM_redshift}, the values of $\alpha_0$, $\beta_0$, $\gamma_0$, and $\sigma_{int_0}$, represent the parameter values at large $t$, which when the shape is flat is represented by the maximum redshift of the sample. This is different than the  $\alpha$, $\beta$, $\gamma$, and $\sigma_{int}$ parameters in GM\&M19 which represent the values at $z=0.0$. Therefore, these two sets of parameters cannot be compared unless the same model is used in both (either a power-law or a Pareto function), which is done and discussed in Section~\ref{sec:DESresults}. 

This Bayesian model regresses the generated values against the observed stellar mass, halo mass, and $\rm m_{gap}$ values simultaneously and self-consistently.  The parameters that model the underlying distributions and their uncertainties are nuisance parameters and thus are marginalized over when we present the posterior distributions.  Each parameter in the Bayesian analysis, along with its prior information is presented in Table \ref{tab:DESbayes}. 

\begin{widetext}
We express the entire posterior as:
\footnotesize
\begin{equation}
\begin{aligned}
p(\alpha_{0},\beta_{0},\gamma_{0},\sigma_{int0},n_{\alpha},n_{\beta},n_{\gamma},n_{\sigma}, x_{i},z_{i},\sigma_{y_i},\sigma_{x_i}, \sigma_{z_i} | x,y,z) \propto & \\ 
& \underbrace{\sum_{i} P(y_{0i}|\alpha_{0},\beta_{0},\gamma_{0},\sigma_{y_i},n_{\alpha},n_{\beta},n_{\gamma},n_{\sigma}, \sigma_{int}, x_{i},z_{i}) ~ P(x_{0i}|x_{i},\sigma_{x_i}) ~ P(z_{0i}|z_{i},\sigma_{z_i})}_{\text{likelihood}} \\ 
&  \underbrace{p(x_i) ~ p(z_{i}) ~  p(\sigma_{x_i}) ~ p(\sigma_{y_i}) ~ p(\sigma_{z_i}) ~ p(\alpha) ~ p(\beta) ~ p(\gamma) ~ p(\sigma_{int}) ~ p(n_{\alpha},n_{\beta},n_{\gamma},n_{\sigma})}_{\text{priors}} 
\end{aligned}
\label{eq:DESposterior}
\end{equation}
\normalsize
\end{widetext}
where each $i^{th}$ cluster is a component in the summed log likelihood.

This is a {\it hierarchical Bayes model} because the priors on the true halo masses ($x_{i}$) and M14 values ($z_{i}$) depend on models themselves (the observed halo mass and M14 distributions).

\section{Results}
\label{sec:DESresults}

\begin{figure*}
    \centering
    \includegraphics[width=18cm]{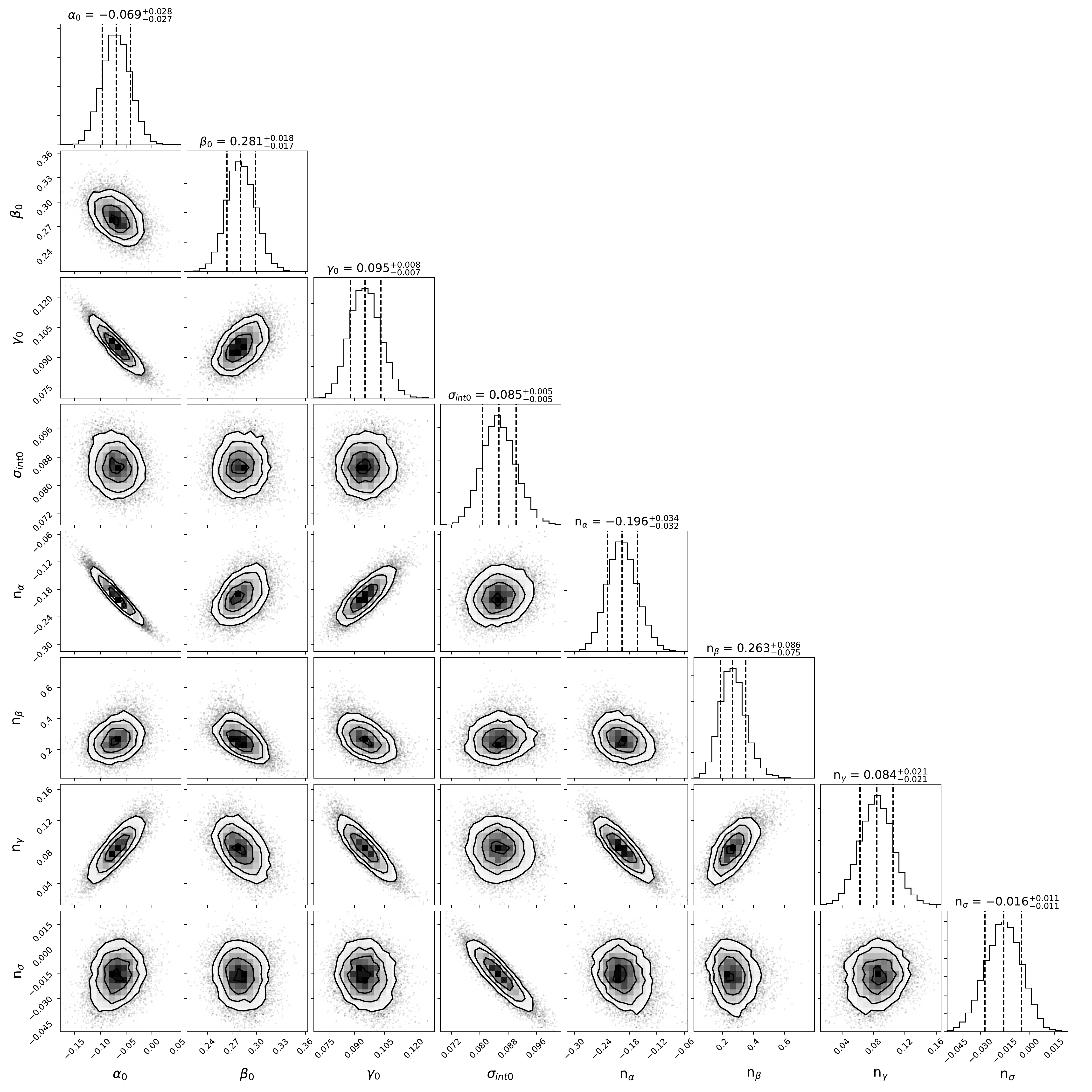}
    \caption{SMHM-M14 parameter posteriors from Equation~\ref{eq:DESSMHM_redshift}. The posterior distribution for $\alpha_{0}$, $\beta_{0}$, $\gamma_{0}$, $\sigma_{int0}$, $n_{\alpha}$, $n_{\beta}$, $n_{\gamma}$, and $n_{\sigma}$.  As in GM$\&$M18, we see that $\gamma$ is significantly non-zero and $\sigma_{int0}$ is less than 0.1 dex.  However, we note that as a result of the modified redshift evolution form given by Equation~\ref{eq:DESSMHM_redshift}, the values are not directly comparable to the results from GM$\&$M19.  Instead, the posteriors for $\alpha_{0}$, $\beta_{0}$, $\gamma_{0}$, and $\sigma_{int0}$ represent the values that these parameters asymptote to.  To see the values at the redshifts measured in our study, see Figure~\ref{fig:DESSMHM_binned}. Using this model, $n_\alpha$ is 5.8$\sigma$ from 0.0,  n$_\beta$ is 3.5$\sigma$ from 0.0, and $n_\gamma$ is 4.0$\sigma$ from 0.0.}
    \label{fig:DESbayesian_redshift_Alpha}
\end{figure*}

We evaluate the strength of the redshift evolution in the SMHM-M14 relation using our previously described MCMC model (Section~\ref{sec:DESmodel}), Bayesian formalism, and linear SMHM relation (Equation~\ref{eq:DESSMHM_redshift}).  For this analysis, we have run the MCMC chains to convergence by examining the parameter autocorrelation functions.  We run our analysis for 10 million steps with a burn in of 2 million steps.  The triangle plot, Figure~\ref{fig:DESbayesian_redshift_Alpha}, shows the 1D and 2D posterior distributions for the eight SMHM relation parameters, $\alpha_{0}$, $\beta_{0}$, $\gamma_{0}$, $\sigma_{int0}$, $n_{\alpha}$, $n_{\beta}$, $n_{\gamma}$, and $n_{\sigma}$. 

A negative (positive) value for the evolution parameters (the $n$'s) indicates that the parameter itself ($\alpha(t)$, $\beta(t)$, $\gamma$(t), $\sigma_{int}(t)$) is growing (shrinking) with increasing lookback time.  The 1D marginalized posteriors in Figure~\ref{fig:DESbayesian_redshift_Alpha} and given in row 3 of Table~\ref{tab:DESbayes} illustrate the evidence for evolution in the SMHM relation in the offset ($n_{\alpha}$), slope ($n_{\beta}$), and stretch ($n_{\gamma}$).  We find no evidence for evolution in the intrinsic scatter ($n_{\sigma_{int}}$), which is small, well within 2$\sigma$ of 0.0,  and consistent with prior results presented in GM$\&$M18 and GM$\&$M19. 

The parameter fits and their errors are provided in Table \ref{tab:DESredshift_ev}. That table starts with constraints from a revised analysis of the data used in GM\&M19. Recall that GM\&M19 used a simple power-law evolution model and 100kpc apertures for the BCG stellar masses.  In addition, the GM\&M19 data only extend to $z=0.3$. Notable differences between GM\&M19 and this work include the use of the Pareto function to describe the evolution, smaller 50kpc aperture stellar masses, and higher redshift data out to $z=0.6$. Therefore, we re-analyze the GM\&M19 data using this new model (Equation \ref{eq:DESSMHM_redshift}) on that original data set, as well as with the new model and using 50kpc apertures for the SDSS BCG stellar masses.

The first row can be compared to the original GM\&M19 discovery of evolution in the slope of the SMHM relation, which was reported at the 3.5$\sigma$ statistical level. Switching from their power law fitting function to the Pareto function, we find that the evolution in the slope is also significantly non-zero, $n_{\beta} = 0.573^{+0.152}_{-0.141}$ or $\sim 3.8\sigma$. We find that by using the Pareto function, the confidence in the detection of the slope evolution has gone up, likely because the Pareto function more closely matches the shape of the data as a function of lookback time.  In the second row, we compare to what happens as we take a smaller physical radius to measure the stellar mass and find no statistically significant differences between the parameter values as a result of using a smaller aperture.

The third row in Table 2 shows the parameter fits for the data described in this paper, which uses DES to extend the analysis to $z=0.6$. The slope evolution is now detected at $3.5\sigma$. Most of the error bars on the parameters have also decreased compared to the analysis on the SDSS data alone. While we have nearly doubled the sample size from GM\&M19, the evolution in $\beta(t)$ is in the late universe where DES does not provide new data. However, the DES data is useful to pin down the amplitude of the flattening of the tail of the Pareto function for the parameters. We note that we have a similar amount of DES data in our higher redshift bins compared to what is present in the SDSS data in GM\&M19 in their highest redshift bin (see Figure \ref{fig:DES_redshift_dist}). Therefore we do not expect a significant drop in the error bars on the inferred parameters when moving from the SDSS-only data set to the combined SDSS and DES data set.

\begin{deluxetable*}{ccccccccc}
    \tablecaption{Posterior Distribution Results with Lookback Time Evolution}
    \tablecolumns{9}
    \tablewidth{0pt}
    \tablehead{ \colhead{Sample} &
    \colhead{$\alpha_0$} &
    \colhead{$\beta_0$} &
    \colhead{$\gamma_0$} &
    \colhead{$\sigma_{int0}$} &
    \colhead{$n_{\alpha}$} &
    \colhead{$n_{\beta}$} &
    \colhead{$n_{\gamma}$} &
    \colhead{$n_{\sigma}$}}
\startdata
 GM$\&$M19\tablenotemark{a} & $-0.325 \substack{+ 0.030 \\ -0.029}$  & 0.224 $\substack{+0.020 \\-0.021 }$ &0.142 $\substack{+0.016\\-0.015}$  & 0.078 $\substack{+0.010 \\-0.009}$ & 0.094 $\substack{+0.068\\-0.061}$ & 0.573$\substack{+0.152 \\ -0.141}$ &0.018 $\substack{+0.026 \\-0.025 }$ & 0.015 $\substack{ +0.018\\ -0.017}$  \\
 GM$\&$M19\tablenotemark{b} & -0.267 $\substack{+0.032 \\ -0.028}$  &0.268 $\substack{+0.024 \\ -0.023}$ & 0.155$\pm$0.016 & 0.081 $\pm$ 0.009& 0.037 $\substack{+0.056 \\ -0.057}$ &0.307 $\substack{+0.105 \\ -0.095}$ &0.003 $\substack{+0.027 \\ -0.025}$ & 0.002 $\substack{+0.016 \\ -0.016}$  \\
 This paper & -0.069 $\substack{+0.028 \\ -0.027}$  &0.281 $\substack{+0.018 \\ -0.017}$ & 0.095 $\substack{+0.008 \\ -0.007}$ & 0.085 $\pm $ 0.005& -0.196 $\substack{+0.034 \\ -0.032}$ &0.263 $\substack{+0.086 \\ -0.075}$ &0.084 $\pm$ 0.021 & -0.016 $\pm$ 0.011 \\ 
 w/o M14  & 0.154 $\substack{+0.015 \\ -0.013}$  & 0.241 $\substack{+0.030 \\-0.027 }$ & -- & 0.098 $\pm$ 0.005 & -0.199 $\pm$ 0.029 & -0.026 $\substack{+0.057 \\ -0.055}$ & -- & 0.013 $\substack{+0.013 \\ -0.012}$  \\
 z $>$ 0.09\tablenotemark{c}  & 0.284 $\substack{+0.086 \\ -0.113}$  & 0.267 $\substack{+0.025 \\-0.023 }$ & 0.076 $\substack{+0.07 \\ -0.006}$ & 0.084 $\pm$ 0.006 & -0.705 $\substack{+0.118 \\ -0.077}$ & 0.260 $\substack{+0.080 \\ -0.079}$ & 0.149 $\substack{+0.016 \\ -0.021}$ & -0.014 $\pm$ 0.016  \\
\enddata
\tablenotetext{a}{The same data from GM\&M19 was re-analyzed using the model from this paper (e.g., equation \ref{eq:DESposterior}) and the original 100kpc apertures.}
\tablenotetext{b}{The same data from GM\&M19 was re-analyzed using the model from this paper (e.g., equation \ref{eq:DESposterior}) and 50kpc apertures for a fair comparison to the results from the data in this paper.}
\tablenotetext{c}{All data in this analysis, except the lowest redshift bin.  We note that while the posterior results differ, when plotted in the redshift range of interest, we find 1$\sigma$ agreement as shown in Figure~\ref{fig:DESSMHM_binned}.}
\tablecomments{Equation \ref{eq:DESSMHM_redshift} parameter fits for the SDSS data in GM\&M19 reference ($z<0.3$) compared to the fits in his paper which use DES data to extend the analysis to $z=0.6$.}
\label{tab:DESredshift_ev}
\end{deluxetable*}

If the parameters $\alpha(t)$, $\beta(t)$, $\gamma(t)$ and $\sigma_{int}(t)$ are evolving between $z=0.3$ and $z=0.6$ , we would expect to see differences in the zero points  $\alpha_0$, $\beta_0$, $\gamma_0$ and $\sigma_{int0}$ between the second and third rows because these parameters represent the value after the Pareto function flattens out to a constant at the upper limit of the redshift traced by the data, which is deeper for DES than for SDSS. We do not detect any changes in $\beta_0$ or $\sigma_{int0}$ after we extend the analysis to $z=0.6$ using the DES data.  However, we do find that $\alpha_0$ is significantly higher and $\gamma_0$ is significantly lower as we extend the data from $z=0.3$ to $z=0.6$. In fact, we detect evolution in the offset $\alpha$ at 5.8$\sigma$ and evolution in the stretch $\gamma$ at 4.0$\sigma$.

We also consider what would happen if we excluded $\rm m_{gap}$ from our model by dropping $z_i$ in Equation \ref{eq:DESSMHM_redshift}. We find that the significance of evolution of the slope drops from $n_{\beta} = 0.263^{+0.086}_{-0.075}$ to being statistically consistent with zero (fourth row in Table 2). Therefore, as originally noted in GM\&M19, the detection of the evolution of the slope of the linear SMHM relation requires the use of $\rm m_{gap}$ in the analysis.

It is interesting that the offset $\alpha$ still shows statistically significant evolution when we ignore $\rm m_{gap}$ in the analysis. We note that in equation \ref{eq:DESSMHM_redshift}, the offset parameter ($\alpha_0$) is not a direct measure of the amplitude of the SMHM. Even when incorporating $n_{\alpha}$, the first term $\alpha(t)$ does not quantify the amplitude of the SMHM (because of the inclusion of $\rm m_{gap}$).  However, without M14, $\alpha(t)$ is simply the overall amplitude of the SMHM relation as a function of lookback time, which is characterized by $\alpha_0$ (at the redshift limit of the data) and $n_{\alpha}$. Thus, when we ignore $\rm m_{gap}$ in our analysis, it appears that we are detecting significant evolution in the amplitude of the SMHM to $z=0.6$. However, assuming that $\alpha(t)$ traces the evolution in the amplitude, the sign on the evolution $n_{\alpha}$ would imply that BCGs are getting more massive as we look back in time. This of course cannot be the case, and we explain this and how to best interpret the observed evolution of $\alpha(t)$ in the next subsection.

\begin{figure}
\subfigure{\includegraphics[scale=0.50]{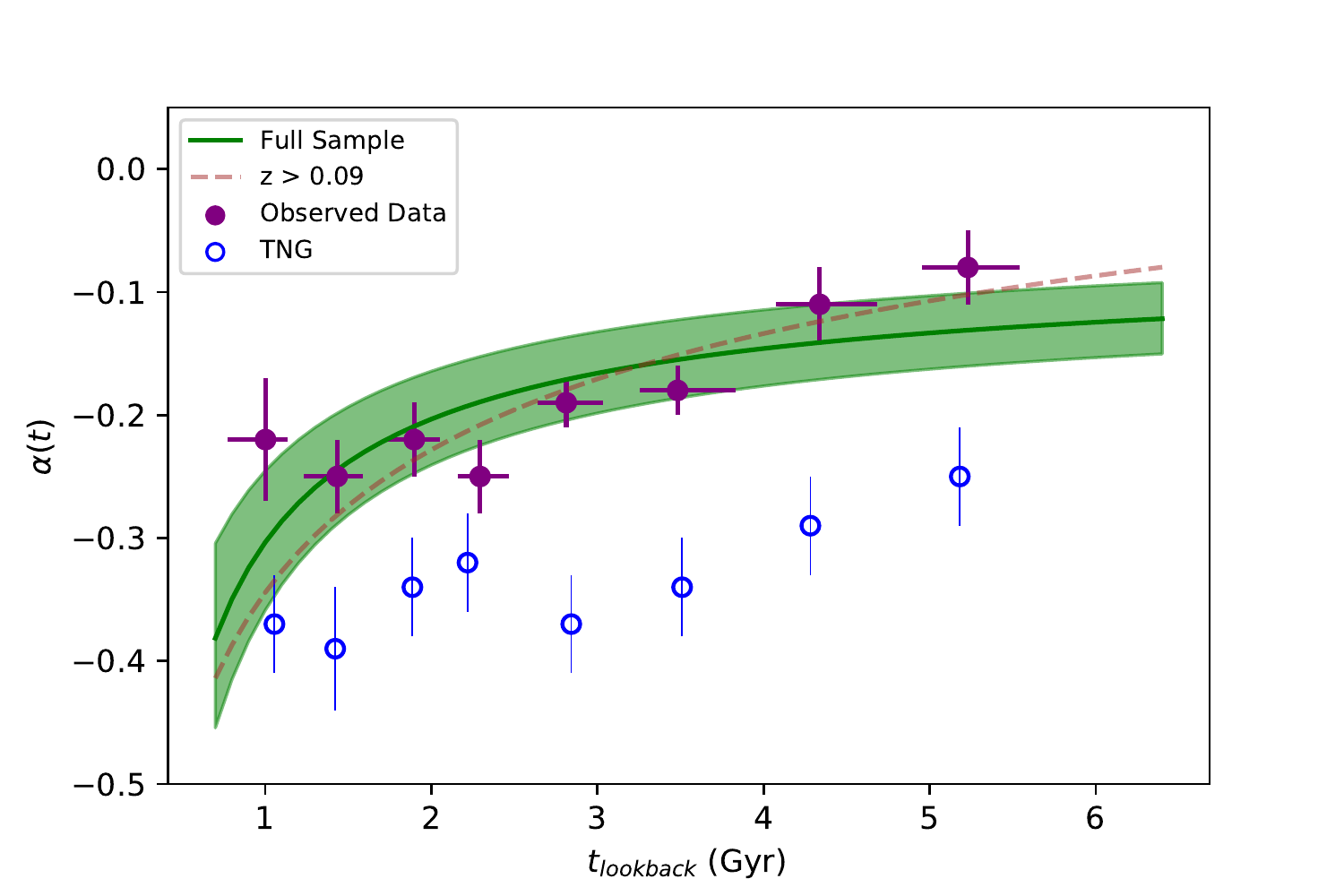}}
\label{fig:DESalpha_redshift_A}
\vspace{-0.8cm}
\subfigure{\includegraphics[scale=0.50]{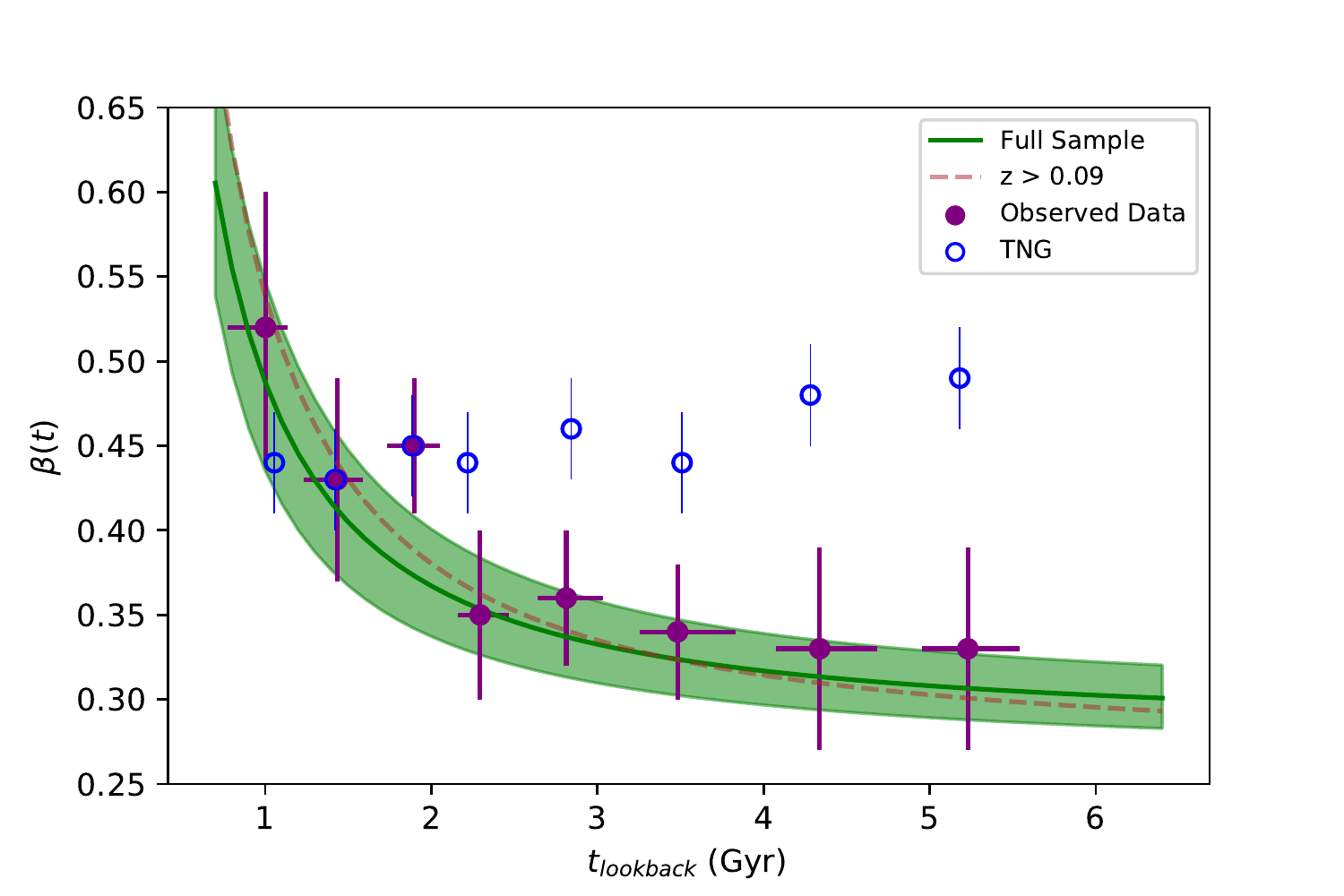}}
\label{fig:DESbeta_redshift_A}
\vspace{-0.8cm}
\subfigure{\includegraphics[scale=0.50]{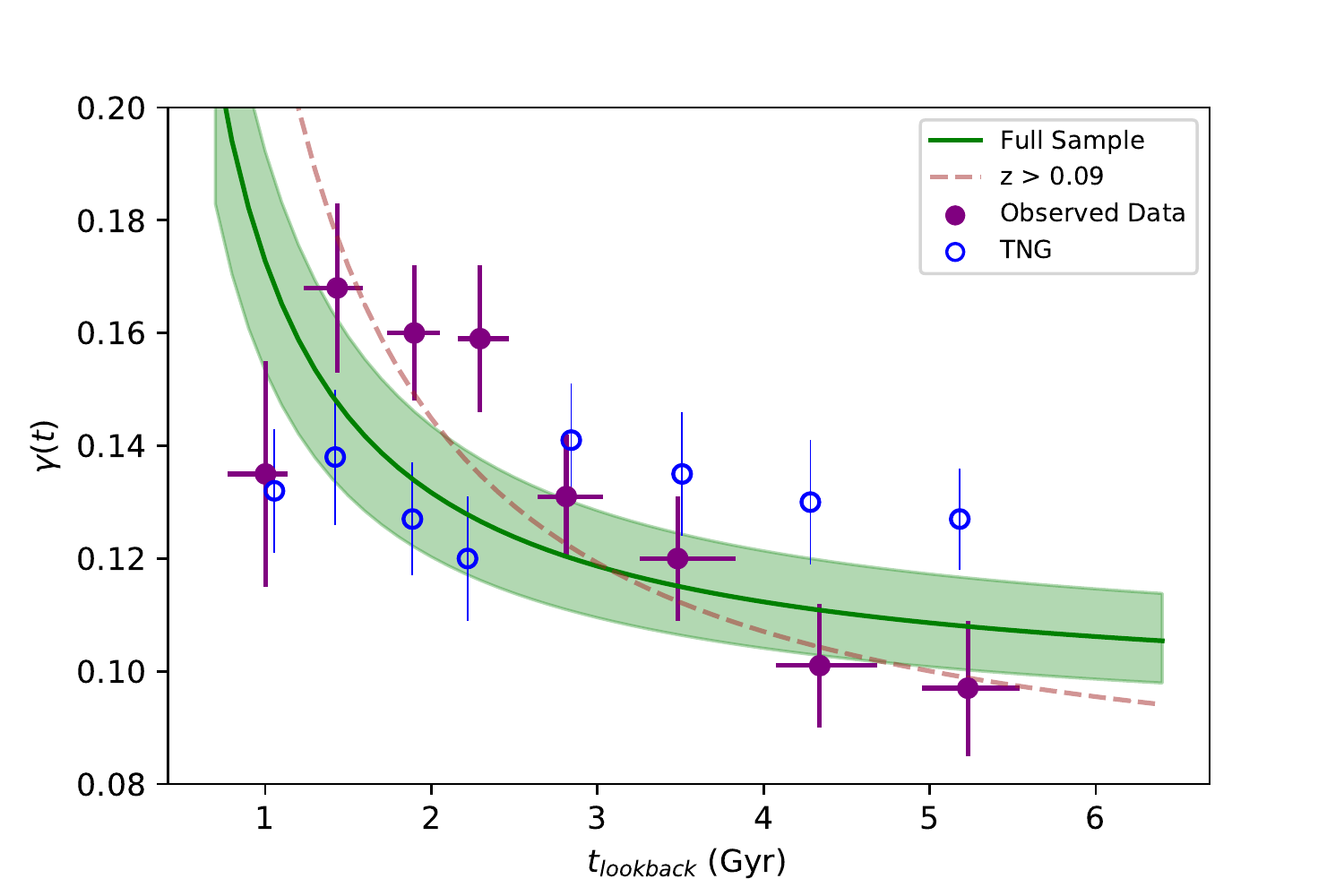}}
\label{fig:DESgamma_redshift_A}
\vspace{-0.8cm}
\subfigure{\includegraphics[scale=0.50]{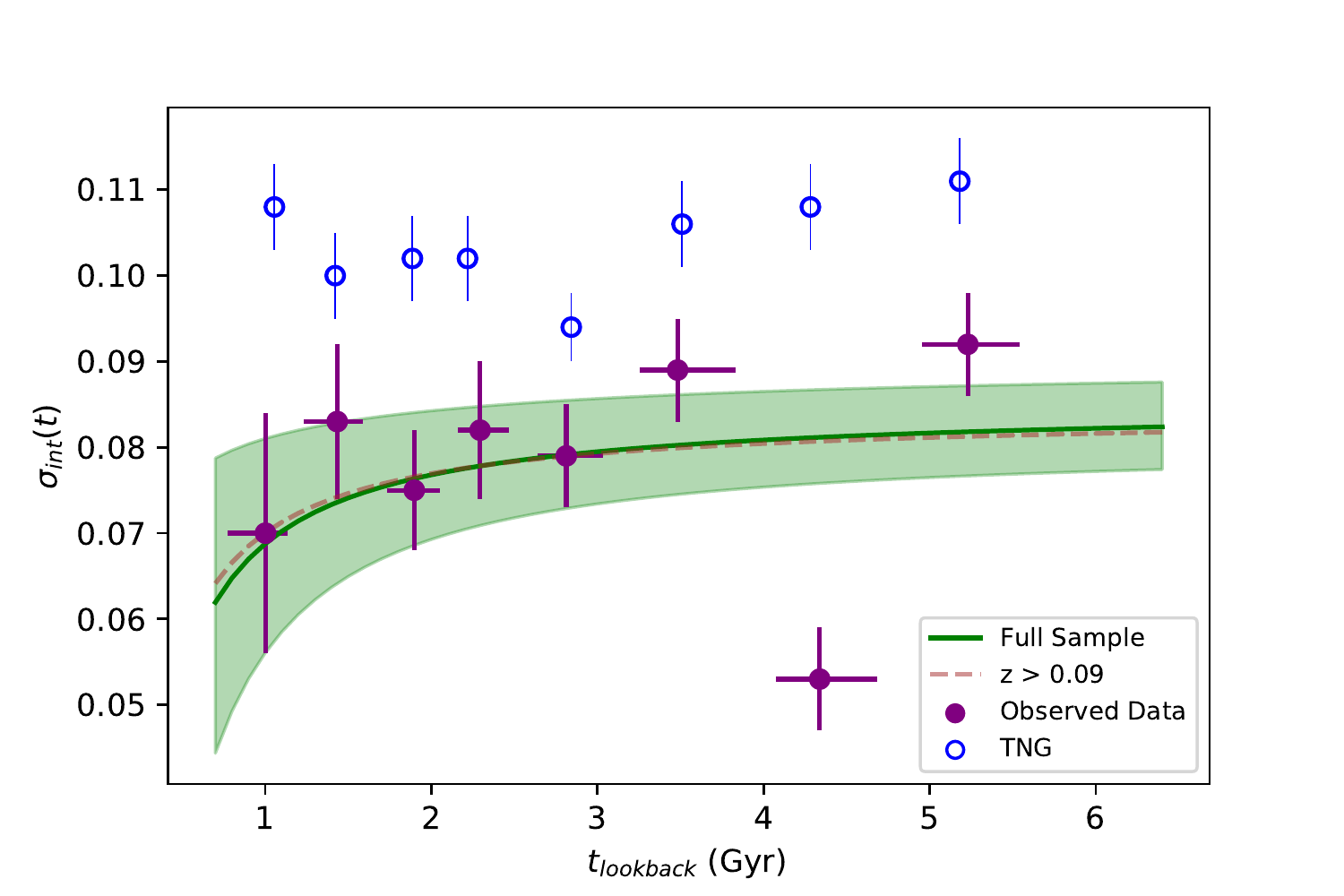}}
\label{fig:DESsigma_redshift_A}
\caption{The effective offset, slope, stretch and intrinsic scatter from Equation \ref{eq:DESSMHM_redshift} as a function of lookback time. The green line represent the result of the fit to the full equation using all of the data. The green error bands are the total error in each parameter as a function of lookback time.  The brown dashed line represents the median posterior when we fit our model without the lowest redshift data bin ($z <$ 0.09). The points represent the redshift binned data when the evolution parameters (e.g., n$_{\alpha}$) are fixed to zero. The error bars contain the middle 67\% of the 1D marginalized posterior. The SDSS and DES clusters are shown in purple and the simulation-based TNG300-1 data in blue.}
\label{fig:DESSMHM_binned}
\end{figure}

\subsection{Comparison to Binned Results}
\label{subsec:DESbin_comp}
By using Equation~\ref{eq:DESSMHM_redshift}, we can characterize the evolution of the SMHM relation through the parameters n$_{\alpha}$, n$_{\beta}$, n$_{\gamma}$, and n$_{\sigma_{int}}$ simultaneously over the full redshift parameter space. However, we can also apply a strong prior and set those parameters to zero, which reverts Equation \ref{eq:DESSMHM_redshift} to Equation \ref{eq:SMHMsimple}. We can then infer $\alpha(t=t_{med})$, $\beta(t=t_{med})$, $\gamma(t=t_{med})$ and $\sigma_{int}(t=t_{med})$ on data separated into discrete redshift bins shown in Figure \ref{fig:DES_redshift_dist} and where $t=t_{med}$ is the median lookback time of the BCGs in each predefined bin. This allows us to make a direct comparison between the fit and a timed step evolution of the SMHM relation. In the binned analysis we assume no evolution within the upper and lower limits on the redshift, which is likely to be true if the time intervals are small enough. We note that the full analysis of Equation~\ref{eq:DESSMHM_redshift} is the correct statistical analysis, since it does not require that assumption and because it does not require a somewhat arbitrary choice of binning. However, the binned analysis provides a good cross check.

In Figure \ref{fig:DESSMHM_binned}, we compare the SMHM parameter values from Equation~\ref{eq:DESSMHM_redshift} with the evolution parameters fixed to zero (purple dots) against the fully evolving parameters for the offset $\alpha(t)$, slope $\beta(t)$, stretch $\gamma(t)$, and intrinsic scatter $\sigma_{int}(t)$ (green line and green band). The purple error bars are the 1$\sigma$ error bars on the binned posteriors. The green error band incorporates the error on the parameter zero points (e.g., $\alpha_0$, $\beta_0$, $\gamma_0$, and $\sigma_{int0}$) as well as the error on the corresponding evolution component ($n_{\alpha}$, $n_{\beta}$, $n_{\gamma}$ and $n_{\sigma_{int}}$).

We find good agreement between the two separate analyses: binned and unbinned.  We first note the evolution in the slope parameter $\beta(t)$ of the SMHM, which is quantified via the Pareto function as $n_\beta = 0.263^{+0.086}_{-0.075}$ and is evident in Figure \ref{fig:DESSMHM_binned}. The Pareto function does a good job capturing the shape of the evolution, which is changing fast at low redshift as originally reported in GM$\&$M19. The slope of the SMHM relation becomes roughly constant beyond a lookback time of 3Gyrs, corresponding to $z$=0.245.  However, in recent times, we clearly identify a steepening of the slope of the SMHM relation for massive clusters.

Figure~\ref{fig:DESSMHM_binned} shows no evidence for evolution in the intrinsic scatter looking back 6 Gyrs ($z=0.6$). The value of the scatter is $\sigma_{int0} = 0.085 \pm{0.005}$. This is the same low value for intrinsic scatter found in GM\&M18 and GM\&M19, except extending to $z=0.6$. We note that an outlier at a lookback time of $\sim 4.3$Gyr exists in the binned analysis.  We are unable to explain this feature of the data, which is an a bin containing exclusively DES data.

In contrast to the intrinsic scatter, Figure~\ref{fig:DESSMHM_binned} shows statistically significant evidence for evolution in both the offset and stretch parameter over the last 6 Gyrs.  Like for the slope, the Pareto function captures the shape of the observed binned evolution, which shows a gradual increase (for $\alpha$) or decrease (for $\gamma$).  Thus, we are clearly identifying an increase in the offset and a decrease in the stretch parameter at higher redshifts.

\begin{figure}
    \centering
    \includegraphics[width=10cm]{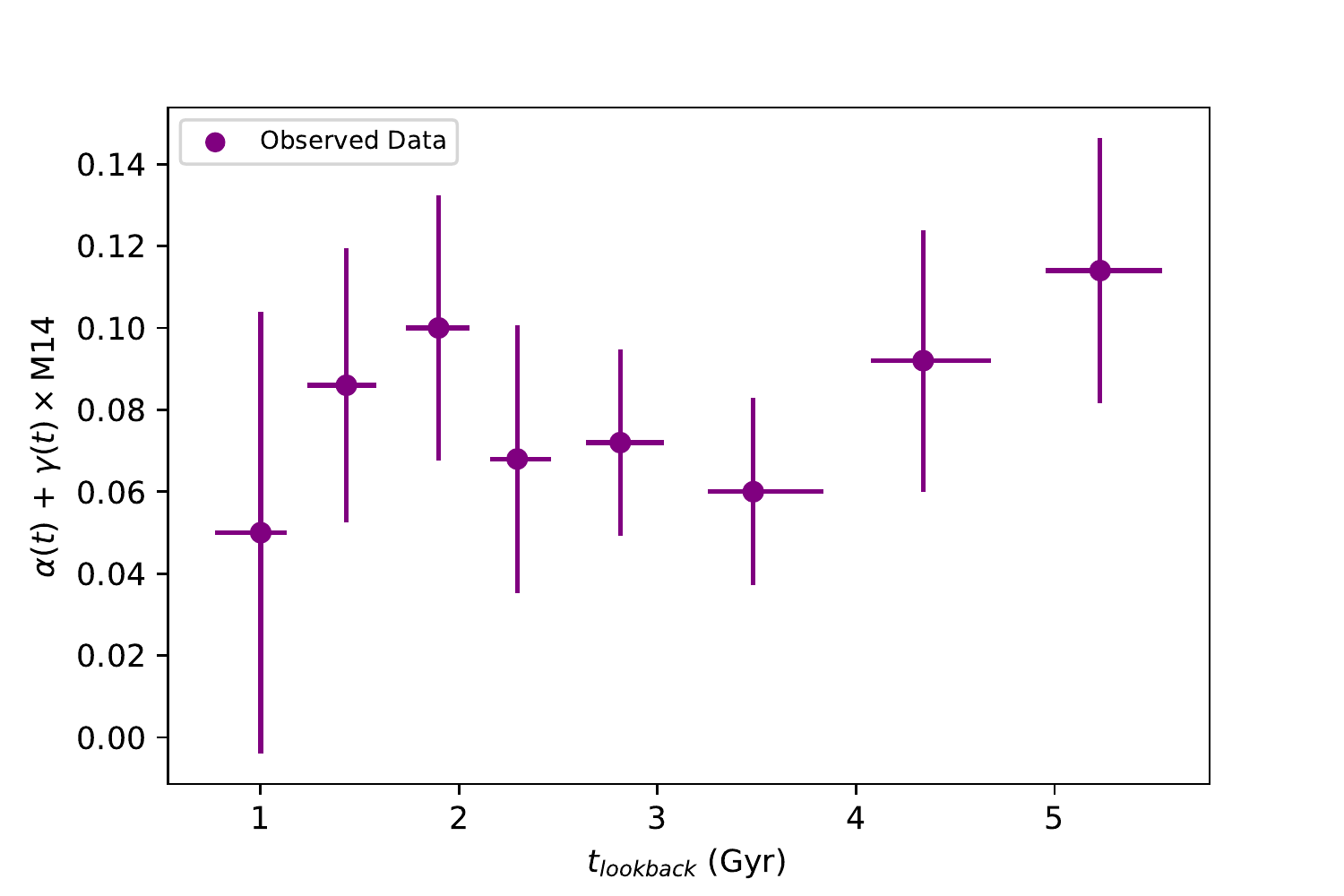}
    \caption{The combined offset of $\alpha(t) + \gamma(t) \times$M14.  This is for when M14=2.0.  This figure illustrates that the combined value of these parameters does not evolve. This indicates that we are not seeing any evolution in the overall amplitude of the SMHM relation, as characterized by the combination of the offset term ($\alpha(t)$) and the stretch term ($\gamma(t)$).}
    \label{fig:DESalpha_gamma}
\end{figure}

As discussed at the end of the previous section, despite detecting evolution in the offset, $\alpha$, we are not actually detecting evolution in the overall amplitude (or median stellar mass at fixed halo mass and $\rm m_{gap}$) for the SMHM relation.  As we show in Figure \ref{fig:DESalpha_gamma}, we plot the combination of the offset and stretch terms, given mathematically as $\alpha + \gamma \times M14$, as a function of lookback time and detect no discernible evolution in the amplitude of the SMHM to $z=0.6$ for a fixed M14 value of 2.0.

\subsection{Statistical Correlations}
In the following subsections, we address correlations in the inferred parameters as well as in the observables. 

\subsubsection{Parameter Correlations}
Figure~\ref{fig:DESbayesian_redshift_Alpha} shows some interesting structure in the 2D posteriors. Besides the obvious correlation between the parameters and their corresponding evolution (i.e., $\alpha$ and $n_{\alpha}$), there is also a weak correlation between the slope and the offset. We note that this correlation has been minimized by re-centering the data using a pivot point selected to be the midway value of the extreme values of the observables.

More importantly, we note the correlation between the offset and the stretch parameters ($\alpha$ and $\gamma$) intertwined with their evolution parameters ($n_{\alpha}$ and $n_{\gamma}$). This was also seen in GM$\&$M19, however that analysis lacked the redshift depth to study the consequences of this correlation. In this work, we have enough data over a large enough lookback time to bin the data beyond where the evolution of the slope flattens.
 
\begin{figure}
    \centering
    \includegraphics[width=10cm]{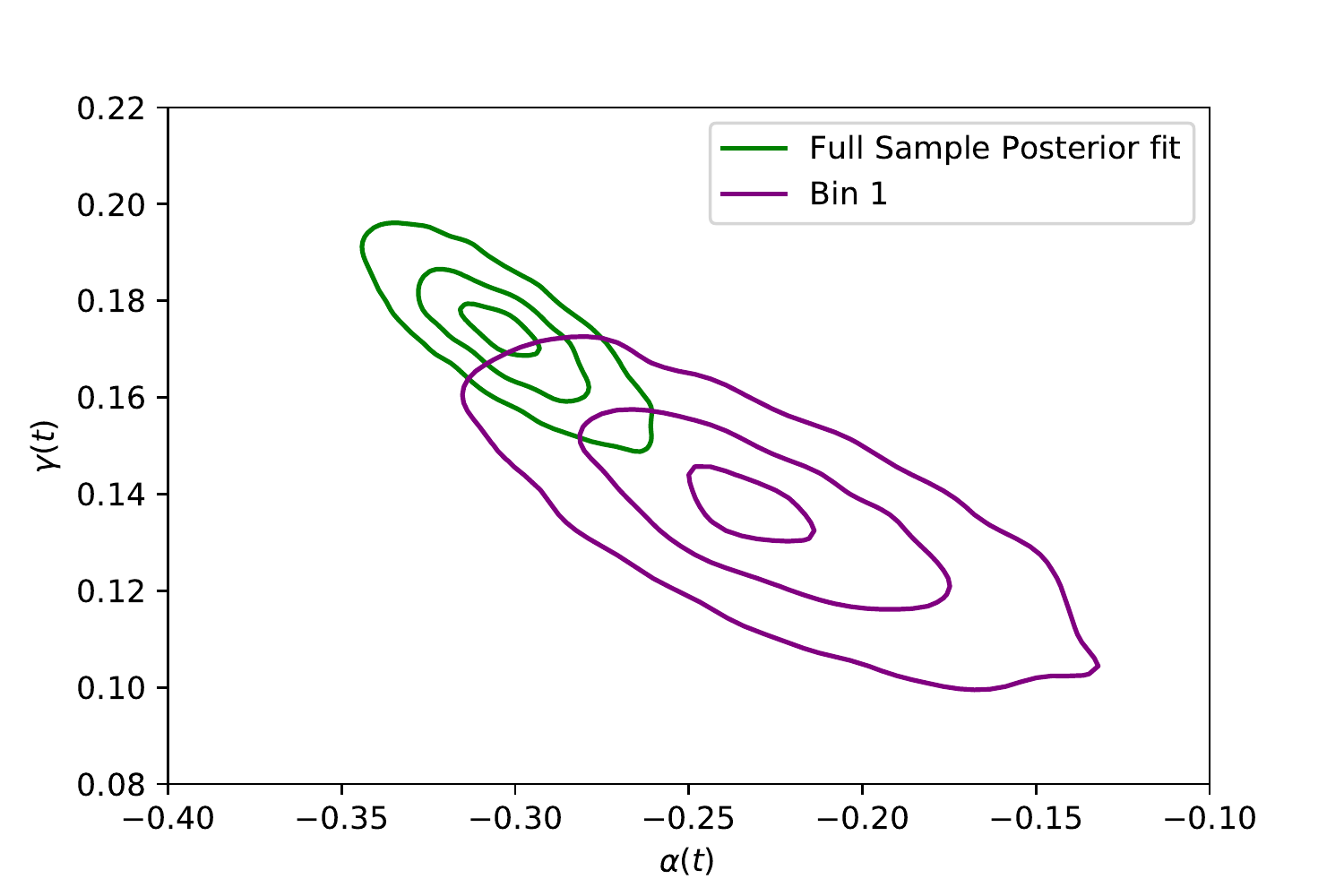}
    \caption{For the first, lowest redshift, bin, we show the contours representative of the 2D posterior distribution for $\alpha$ and $\gamma$ in purple.  A strong covariance exists between these parameters.  In green, we overlay the posterior distribution for these same parameters as estimated from Figure~\ref{fig:DESSMHM_binned} for the median lookback time of bin 1.}
    \label{fig:Bin1alpha_gamma}
\end{figure}

In Figure~\ref{fig:DESSMHM_binned}, we notice that $\alpha(t)$ and $\gamma(t)$ in the lowest redshift bin are $\sim$ 1$\sigma$ low ($\alpha(t)$) and $\sim$ 1$\sigma$ high ($\gamma(t)$) when compared to unbinned fits. While some other bins have similar differences, this is the only bin where the binned values do not follow the general trend displayed by the green posterior distributions (even though the measured values are within 1$\sigma$.  We explain this discrepancy via the covariance between $\alpha$ and $\gamma$, evident in Figure~\ref{fig:DESbayesian_redshift_Alpha}. As the parameter $\alpha$ scatters low in the MCMC sampling, $\gamma$ scatters high. There is a clear degeneracy in these two parameters.  We show this degeneracy just for clusters in the lowest-redshift binned analysis in Figure~\ref{fig:Bin1alpha_gamma}. We overplot the 2D posterior distribution between $\alpha(t)$ and $\gamma(t)$ for the low-redshift bin (in green) and the total posterior (in purple).  
Figure~\ref{fig:Bin1alpha_gamma} highlights that we see that the 2D posterior error ellipses associated with the single-bin analysis overlap with those measured based on the entire sample.  However, we note that there is likely a weak covariance between the two sets of 2 dimensional posteriors that may be responsible for part of this agreement.  

Figure~\ref{fig:Bin1alpha_gamma} exemplifies why one would may want to avoid binning in this type of SMHM analysis, since unaccounted parameter covariances can lead to incorrect fits to the evolving SMHM relation. Our fully unbinned analysis and our hierarchical Bayesian formalism allow for these covariances to naturally be accounted for in the fitted parameters and their marginalized posteriors.   However, we note that despite this reservation, the results from the binned analysis are largely consistent with our evolution analysis.  For an additional test, we fit our model to the data after excluding those clusters in the lowest redshift bin ($z < $0.09). We show the median posterior for $\alpha(t)$, $\beta(t)$, $\gamma(t)$, and $\sigma_{int}(t)$ as the brown dashed line in Figure~\ref{fig:DESsigma_redshift_A}.  The entire posterior is given in line 4 of Table~\ref{tab:DESbayes}. While the median posterior values differ (likely due to the covariances between the parameters and their evolution, when plotted as a function of lookback time, there are no significant differences.  We do note that as evident from Figure~\ref{fig:Bin1alpha_gamma}, a higher stretch parameter at low redshift is still preferred.

\subsubsection{Data correlations}
We make some assumptions in Equation \ref{eq:DESSMHM_redshift} and in Section~\ref{sec:DES}. Primarily, we assume that the observables (stellar mass, halos mass, and $\rm m_{gap}$) are independent observables. If our data were strongly correlated to each other in some complicated away (or to some latent parameter), we would need to quantify those correlations and their impact on the fitted SMHM parameters. Our main concern is that unlike GM$\&$M18, we use richness as a proxy for halo mass and a correlation could exist between richness and either $\rm m_{gap}$ or the BCG stellar masses that would affect our conclusions.

\citet{hea13} reported a correlation between M12 and the cluster richness at fixed halo mass. While they do not quantify the correlation, they suggest that there is evidence that having a large $\rm m_{gap}$ is correlated with being under-rich at a given halo mass.  At fixed X-ray luminosity (as a proxy for halo mass), \citet{Erfanianfar19} report a weak and positive correlation between the cluster richness and the stellar mass of BCGs (Pearson correlation coefficient $\sim 0.2$). \citet{Furnell18} used dynamical masses to find a similar weakly positive correlation between richness and BCG stellar mass (Spearman rank correlation coefficient $r_s = 0.137$). On the other hand, \citet{Farahi20} used the Illustris TNG simulations to find a moderate {\it anti-correlation} between richness and BCG stellar mass at fixed halo mass (Pearson correlation coefficient $\sim -0.4$). None of these reported correlations are strong or consistent. Compounding the issue is that the richnesses, halo masses, stellar masses, and $\rm m_{gap}$ measurements are not homogeneously measured in either the data or the simulations.

Given the above information, and the fact that we do not have other halo mass proxies like X-ray luminosity, weak lensing, or dynamical masses for our clusters, there is little we can do in terms of a precision exploration of the data correlations. However, we can calculate the Pearson correlation coefficient between our richness inferred halo masses and $\rm m_{gap}$ at fixed BCG stellar mass. If we fix the stellar mass to within $11.4 < log_{10}(M_{\star}$ /$(M_{\odot}/h^2)) < 11.6$, the stellar mass range that allows us to measure the correlation across the entire redshift range, we find a moderate anti-correlation of $\sim -0.4$.

The statistical significance of correlation coefficients is not well defined. Most in the literature use some form of jacknife sampling \citep[e.g.,][]{Erfanianfar19}.  However, here, we use a Bayesian-like approach where we apply Equation~\ref{eq:SMHMsimple} to forward model our data using uncertainties given in section \ref{subsec:DESuncertainties}. We can then apply a correlation between $\rm m_{gap}$ and the halo mass before the simulated observational uncertainties are incorporated. We then run 10,000 simulations with and without the correlation and measure the standard deviation on the measured correlation coefficient as well as the probability of the correlation coefficient being observed in a purely non-correlated data set (i.e., a null test). We find that the error on the correlation is $\sim 0.04$ and the probability of a purely randomized data set showing the same level of correlation we find to be $p = 0.001$. We conclude that correlation between M14 and $M_{200m}$ (as inferred by richness) is significant. We note that we find a nearly identical anti-correlation between M14 and $M_{200}$ in the Illustris-TNG sample ($-0.36$).  

We can use this same forward modeling technique to quantify the effect this correlation could be having on the parameters we measure when assuming independence. We note that this is not the same as developing a new statistical model which incorporates correlations between the data, which we reserve for a future effort. Using this Bayesian-like approach, we do however, estimate the impact of this correlation on the slope.  We find that the correlation between M14 and $M_{200}$ results in an increase in the slope by approximately 0.15.  However, we note that because this correlation persists across the redshift space, we do not believe that it impacts our detected redshift evolution, but rather just the measured value of the slope.  Thus, this analysis provides us with a good idea of the level of the effect of the  correlation on the slope, offset, and stretch without introducing one or more new free parameters to the model. We will explore these interesting correlations in a future analysis.

\subsection{Comparison to Illustris TNG300-1}
\label{subsec:TNGcomp}

Figure~\ref{fig:DESSMHM_binned} offers direct comparisons between the observed (SDSS and DES) results and the simulated TNG300-1 measurements.  Such a comparison allows us to understand whether the physical prescriptions built into the TNG300-1 simulated universe yield observations that match those found in the observed universe.  This analysis is designed to yield a fair comparison since for both data sets we subtract off the same median in stellar and halo mass, which is based on the observed SDSS and DES values, allowing the posterior distributions to be directly compared.  In such a comparison, $\alpha$ is related to the median stellar mass (at a halo mass of log$\rm_{10}(M_{halo}$ /$(M_{\odot}/h))=14.65$) at a given $\rm m_{gap}$ (since the $\gamma$ values agree).  We note that for the simulated data, like for the observational data, M$\rm_{halo}$ refers to M$\rm_{200m}$.

The only similar result between the observed and simulated universes is the lack of evolution of $\sigma_{int}$; we detect no evolution in either.  Interestingly, \citet{pil17} detect modest evolution in $\sigma_{int}$, such that from $z$=0.0 to $z$=0.5, the value increases by $\sim$ 0.04 dex.  However, the results presented in \citet{pil17} do not account for $\rm m_{gap}$. In contrast, when we measure the evolution in the SMHM relation without incorporating $\rm m_{gap}$, we do not find this evolution, though the size of our error bars may prevent us from detecting it. 

One of the more significant results from using our approach to measure the 50kpc magnitudes for the TNG300-1 data, is the absence of noticeable evolution in the slope of the SMHM relation for TNG300-1.  In our observed data set, late time growth appears to occur primarily in the last 2 billion years; however, in the TNG300-1 simulation, there is no detectable evolution over the entire time range studied.  However, we note that the absence of redshift evolution in the slope with the TNG300-1 data agrees with \citet{pil17} and \citet{eng20}, who claim no such evolution.  Thus, unlike for observations, where GM$\&$M19 found that the incorporation of m$\rm_{gap}$ led to the detection of evolution, for the TNG300-1 simulation this is not the case.

Another difference between our TNG300-1 and prior measurements from \citet{pil17} and \cite{eng20} is the value of $\beta$.  We measure a value of approximately 0.42 for the slope when the stellar mass is measured within 50kpc when $\rm m_{gap}$ is incorporated and 0.48 when it is not. Our estimate is therefore in agreement with \citet{pil17} who measure the stellar content within 30kpc and find a slope of 0.49 (no error bars reported).  We note our slope is much shallower than that measured in \citet{eng20} and other slopes measured in \citet{pil17}, $\approx$0.70, which are measured using the 2 times the stellar half mass radius, a radius far greater than the 50kpc aperature we use.  Therefore, we can conclude, as shown in GM$\&$M19 that had we used a large aperture ($>=$ 100kpc) to measure the BCG stellar mass and magnitude, then we would likely recover a steeper slope.  One additional note is that both here and in GM$\&$M19, we find that the slope of the SMHM relation is steeper when $\rm m_{gap}$ is incorporated, which serves as evidence that incorporating information about the satellite galaxy population (via $\rm m_{gap}$) yields a steeper slope than the traditional SMHM relation, which agrees with the general conclusions from \citet{tin19}.  However, as shown in Table~\ref{tab:DESsim_comp}, this trend is not shown in the TNG300-1 data.  Instead, we see that the slopes are within 1$\sigma$, which may serve as the first bit of evidence that the BCGs and growth prescriptions in the TNG300-1 simulation are over-dominant.  

The remaining two parameters $\alpha$ and $\gamma$ are also dramatically different.  Unlike in our observational data, there is little evidence of any evolution in $\alpha$ or $\gamma$ out to high redshift.  Given that $\alpha$ and $\gamma$ are covariant, it is unsurprising that if one of these two parameters shows no evolution, the other parameter also shows not evolution, and as discussed in Section~\ref{sec:DESdiscussion} likely related to the growth prescription used in TNG.  Additionally, the values for $\alpha$ also significantly differ.  At first glance, it appears as though the TNG300-1 BCGs are undermassive.  However, that is not the case.  A more valid comparison would be the value of $\alpha + \gamma \times z_{med}$, which would be representative of the median stellar mass of the BCGs at a given halo mass.  This comparison yields that the TNG systems are approximately 0.06 dex overmassive.  Of note, when doing such a comparison, the median M14 values for TNG are approximately 1-1.5 magnitudes greater than the observed values; unlike in the observed universe, low $\rm m_{gap}$ systems (M14$<$1.0) do not exist.  While our measurements suggest that part of this difference is a result of the slightly overmassive BCGs, for such a scenario to occur, it is likely that the merging prescription used in TNG300-1 also results in poorly populated red sequences, such that few intermediate brightness galaxies exist, thus yielding substantially fainter 4th brightest galaxies.      

\section{Discussion}
\label{sec:DESdiscussion}
In GM$\&$M19, we introduced the novel observation of evolution in the slope of the SMHM relation and used that observation to offer insight into the late-time hierarchical growth of BCGs.  As shown here, significantly expanding the parameter space out to higher redshifts/earlier lookback times using DES-redMaPPer data, we reach a much deeper understanding of how BCGs and the clusters that they reside within grow and evolve over the last 6 billion years.  

Currently, there is not a clear consensus between observations, simulations, and models about how BCGs grow over this redshift range.  Using semi-analytic models, researchers have found that at late times $(0.0 < z < 0.5)$ BCG's grow by a factor of $\approx$1.5-2.0 \citep{del07, guo10, shankar15}.  In contrast observations suggest that over this redshift range, much of the growth occurs in the BCG's outermost envelope, incorporating regimes that are often characterized as being part of the ICL \citep{van2010,bur15,hua18,fur21}, which highlights the necessity of looking at the BCG+ICL system jointly.  However, in this work we use the additional information provided via the inclusion of $\rm m_{gap}$ into the SMHM relation to determine physically what growth is occurring in the BCG+ICL system over this redshift range.   

In this work, we extend the redshift evolution of the cluster scale SMHM presented in GM$\&$M19 ($0.03 < z < 0.30$) out to $z_{red}=0.6$.  To briefly summarize our findings, we confirm all key results from GM$\&$M18 and GM$\&$M19: $\rm m_{gap}$ is definitively a latent parameter within the SMHM relation; incorporating $\gamma$ and M14 into the SMHM relation reduces $\sigma_{int}$; and accounting for $\rm m_{gap}$ yields significant evolution in the slope of the SMHM relation over late time.  From this analysis, we for the first time, report evolution in both the $\alpha$ and $\gamma$ parameters, which represent the offset and stretch, respectively.  It is this observed evolution that drives our understanding of how BCG's evolve.

\begin{figure}
    \centering
    \includegraphics[width=9cm]{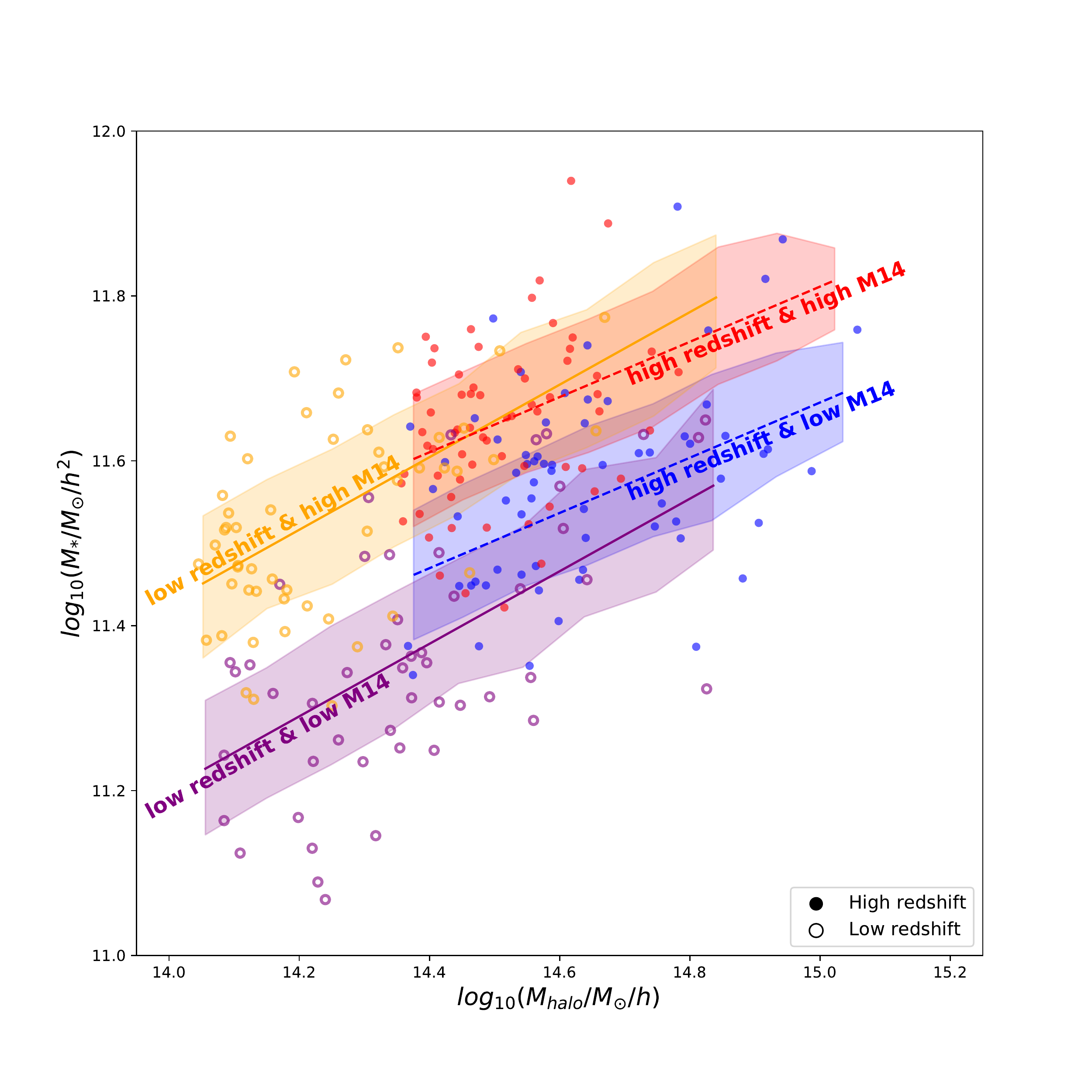}
    \caption{We display two sets of distributions for the the low (unfilled) and high-$z$ (filled) data.  For both, we show two $\rm m_{gap}$ regimes, the upper 80-90\% regime (orange and red) and the lower 10-20\% regime (purple and blue).  The shaded regions represent the posterior distributions from our models.  As shown, for the low redshift data, we see a steeper slope and more pronounced stratification, which results from a larger stretch.}
    \label{fig:DES_stratification}
\end{figure}

To understand how the stellar mass, halo mass, and $\rm m_{gap}$ are changing as a function of lookback time (or redshift), in Figure~\ref{fig:DES_stratification}, we plot the SMHM relation data for a low redshift sample (2nd and 3rd bin) and a high redshift sample (7th and 8th bin).  We note that due to lack of data and the larger difference in parameter values, we do not use the lowest redshift bin.  For each sample, we plot the data in the 10th-20th and 80th-90th percentiles of the $\rm m_{gap}$ distribution.  This is shown by the filled in (high-$z$) and unfilled (low-$z$) data points.  We then overlay the results of the posterior distributions shown in Figure~\ref{fig:DESbayesian_redshift_Alpha} and Figure~\ref{fig:DESSMHM_binned} as the shaded regions.  
Figure~\ref{fig:DES_stratification} highlights a few of our key findings.  First, $\gamma$ is significantly growing as one moves forward in lookback time, as evidenced by how much larger the separation between the two shaded regions are at low-$z$ when compared to high-$z$.  We note that if $\gamma$ were not evolving, the separation between the two high and low M14 bins would not be growing, regardless of the change in halo mass distribution of the data, since $\gamma$ does not vary with M$\rm_{halo}$.  Second, $\beta$ is growing as one moves forward in lookback time.  Third, $\sigma_{int}$, the spread in the data at fixed $\rm m_{gap}$ is unchanged between these two distributions, which supports our measurement that $\sigma_{int}$ is not evolving.  Fourth, the most insightful observation shown here, as highlighted by the regions in the M$\rm_{halo}$ distribution where these data sets overlap (14.4 $\rm < log_{10}(M_{halo}$ /$\rm (M_{\odot}/h)) <$ 14.7), the BCG stellar mass distribution remains the same, and thus the BCG stellar mass within 50kpc is not growing.  This is also supported by the constant value of $\alpha + \gamma \times M14$ given in Figure~\ref{fig:DESalpha_gamma}.

In GM$\&$M19, given the absence of evolution in $\gamma$, we assumed that any growth observed was due entirely to growth in the BCG.  However, as shown in Figure~\ref{fig:DES_stratification} the stellar mass within 50kpc is not growing over this redshift range.  This observation highlights that the driver behind all the evolution we detect and have previously detected is instead $\rm m_{gap}$.  First, with respect to the slope, if the stellar mass is not changing, the only way for the slope to increase would be for the stellar masses of the distribution of clusters that are linked by having similar $\rm m_{gap}$ values to change, as a result of changing $\rm m_{gap}$ values.  For the slope to increase, this would likely be in such a manner that the most massive systems, with the more massive BCGs have $\rm m_{gap}$ values that are growing more efficiently and quickly, likely due to their residing in richer clusters.  

Recall that $\rm m_{gap}$ is the difference in brightness between the BCG and 4th brightest cluster member \citep{dar10} and results from the hierarchical assembly of the BCG (GM$\&$M18), such that we expect clusters characterized by larger $\rm m_{gap}$ values to form earlier). Since the observed evolution results from changes in the $\rm m_{gap}$ distribution, the most insight into what is physically happening can be instead gleaned from the evolution in $\gamma$.  For $\gamma$ to evolve, the $\rm m_{gap}$ distribution must be changing with time.  This is not happening in a manner that changes the BCG stellar mass (within 50kpc).  Therefore, instead, what is likely happening is that mergers between the bright satellite galaxies and the BCG deposit the stellar material at radii beyond 50kpc, what we interpret as the ICL.  Therefore, the outer envelopes contain all recent BCG growth and it is only through the incorporation of $\rm m_{gap}$ that we are able to detect this evolution without measuring the BCG + ICL profile as done in \citet{zha18}.  As a result of this scenario, the separation of clusters with fixed $\rm m_{gap}$ values, what we refer to as our stratification, becomes larger while the stellar mass distribution (within 50kpc) remains fixed.  Therefore, the incorporation of $\rm m_{gap}$ has elucidated that BCGs continue to grow hierarchically in this redshift range, but all of that added stellar material is going directly into the growth of the ICL.  This result is supported observationally by \citet{fur21}, who find evidence of ICL growth over $0.1 < z < 0.5$.      

While the main takeaways of this paper are observational, we do want to comment on what the absence of evolution in TNG300-1 means.  Since we detect no evolution in either the slope, stretch, or offset parameters, clearly the same kind of hierarchical growth prescription is not occurring within the TNG300-1 simulation.  Additionally, the TNG300-1 clusters are characterized by larger $\rm m_{gap}$ measurements.  Therefore, it is possible that the majority of the stellar mass within these BCGs is assembled at earlier times.  Moreover, due to an over efficient merger process, there exists an absence of fainter satellite galaxies in the TNG300-1 simulation, the same population that we observationally find must be responsible for the continued hierarchical assembly of the BCG + ICL systems.  

In this work, we have focused on the late time evolution of the SMHM relation out to $z \sim 0.6$.  As shown here, $\beta$ shows significant late-time evolution, predominately over the redshift range $0.0 < z < 0.15$ and we for the first time detect statistically significant evolution in $\alpha$ and $\gamma$, which has clarified that this evolution is driven by BCG hierarchical growth that is evident not in the stellar mass, but rather within the $\rm m_{gap}$.  We are left with a few paths forward.  If we choose to tighten the constraints further on this late-time evolution, we must either incorporate more large statistically complete samples of low-redshift clusters $z < 0.1$ (there are fewer than 200 SDSS low-$z$ clusters compared to $\sim$1300 DES high-$z$ clusters), which are difficult to obtain or, we can forge ahead to higher redshifts to determine whether these parameters continue to evolve out to $z = 1.0$, using a data set such as the DES-ACT overlap \citep{Hil20} or the DES-SPT overlap \citep{ble15, ble20, hua20}, an approach which faces similar observational and modelling challenges as the results presented here, but presents the opportunity for us to further quantify and better constrain this evolution.  Additionally, given that we have now statistically verified that the stellar mass - m$\rm_{gap}$ trend exists in both observations and state-of-the-art hydrodynamic simulations, although we note that the evolution trends do not match, a key step forward may be to determine the physical meaning of this correlation between stellar mass and $\rm m_{gap}$, what it may inform us about the formation history of the BCG and its host cluster dark matter halo, and quantify how the stellar mass, halo mass, $\rm m_{gap}$ parameter space maps to a cluster's formation redshift.  Lastly, as explored in GM$\&$M19, we should continue to study the BCG light profiles out to large radii of 100kpc and beyond.  Another vital step forward as part of that effort is to take advantage of the ICL measurements done by \citet{zha18} for the DES clusters to determine whether we are able to detect significant growth in the ICL over this redshift range.  Such a result would verify our conclusion that all the recent growth is contained within the ICL and that it's these recent mergers, which change the $\rm m_{gap}$ distribution and yield our detected evolution. 

\acknowledgements
This paper has gone through internal review by the DES collaboration.  JGM would like to thank Emmet Golden-Marx for useful discussions about generating the photometric images used in this paper and Ying Zu for key discussions about the cosmological measurements that were used in this analysis.  JGM acknowledges the support by the National Key Basic Research
and Development Program of China (No. 2018YFA0404504) and the National Science
Foundation of China (No. 11873038, 11890692).

Funding for the DES Projects has been provided by the U.S. Department of Energy, the U.S. National Science Foundation, the Ministry of Science and Education of Spain, the Science and Technology Facilities Council of the United Kingdom, the Higher Education Funding Council for England, the National Center for Supercomputing Applications at the University of Illinois at Urbana-Champaign, the Kavli Institute of Cosmological Physics at the University of Chicago, the Center for Cosmology and Astro-Particle Physics at the Ohio State University, the Mitchell Institute for Fundamental Physics and Astronomy at Texas A\&M University, Financiadora de Estudos e Projetos, Funda{\c c}{\~a}o Carlos Chagas Filho de Amparo {\`a} Pesquisa do Estado do Rio de Janeiro, Conselho Nacional de Desenvolvimento Cient{\'i}fico e Tecnol{\'o}gico and the Minist{\'e}rio da Ci{\^e}ncia, Tecnologia e Inova{\c c}{\~a}o, the Deutsche Forschungsgemeinschaft and the Collaborating Institutions in the Dark Energy Survey. 

The Collaborating Institutions are Argonne National Laboratory, the University of California at Santa Cruz, the University of Cambridge, Centro de Investigaciones Energ{\'e}ticas, Medioambientales y Tecnol{\'o}gicas-Madrid, the University of Chicago, University College London, the DES-Brazil Consortium, the University of Edinburgh, the Eidgen{\"o}ssische Technische Hochschule (ETH) Z{\"u}rich, Fermi National Accelerator Laboratory, the University of Illinois at Urbana-Champaign, the Institut de Ci{\`e}ncies de l'Espai (IEEC/CSIC), the Institut de F{\'i}sica d'Altes Energies, Lawrence Berkeley National Laboratory, the Ludwig-Maximilians Universit{\"a}t M{\"u}nchen and the associated Excellence Cluster Universe, the University of Michigan, NSF's NOIRLab, the University of Nottingham, The Ohio State University, the University of Pennsylvania, the University of Portsmouth, SLAC National Accelerator Laboratory, Stanford University, the University of Sussex, Texas A\&M University, and the OzDES Membership Consortium.

Based in part on observations at Cerro Tololo Inter-American Observatory at NSF's NOIRLab (NOIRLab Prop. ID 2012B-0001; PI: J. Frieman), which is managed by the Association of Universities for Research in Astronomy (AURA) under a cooperative agreement with the National Science Foundation.

The DES data management system is supported by the National Science Foundation under Grant Numbers AST-1138766 and AST-1536171. The DES participants from Spanish institutions are partially supported by MICINN under grants ESP2017-89838, PGC2018-094773, PGC2018-102021, SEV-2016-0588, SEV-2016-0597, and MDM-2015-0509, some of which include ERDF funds from the European Union. IFAE is partially funded by the CERCA program of the Generalitat de Catalunya.  Research leading to these results has received funding from the European Research Council under the European Union's Seventh Framework Program (FP7/2007-2013) including ERC grant agreements 240672, 291329, and 306478.  We  acknowledge support from the Brazilian Instituto Nacional de Ci\^enciae Tecnologia (INCT) do e-Universo (CNPq grant 465376/2014-2).

This manuscript has been authored by Fermi Research Alliance, LLC under Contract No. DE-AC02-07CH11359 with the U.S. Department of Energy, Office of Science, Office of High Energy Physics.

%\bibliography{SMHM_DES}

\begin{appendix}
In this appendix we provide the fits to the parameters in the binned analysis of Figure~\ref{fig:DESSMHM_binned} (including both the observational and simulated data), along with the fits for the data subsets that were used to calibrate the uncertainties on the observables.
\begin{deluxetable*}{cccccccccc}
    \tablecaption{Posterior Distribution Results}
    \tablecolumns{10}
    \tablewidth{0pt}
    \tablehead{ \colhead{Data} & 
    \colhead{${z}_{min}$} & 
    \colhead{${z}_{max}$} &
    \colhead{${z}_{med}$} &
    \colhead{$log_{10}$($M_{halo_{min}}$)} &
    \colhead{$n_{clusters}$} &
    \colhead{$\alpha$($z=z_{med}$)} &
    \colhead{$\beta$($z=z_{med}$)} &
    \colhead{$\gamma$($z=z_{med}$)} &
    \colhead{$\sigma_{int}$($z=z_{med}$)}}
\startdata
GM$\&$M bin 4 (100kpc) & 0.208 & 0.300 & 0.247 & 14.39 & 210 &$-0.34 \pm 0.03$ & $0.32 \pm 0.06$ & $0.150 \pm 0.013$ & $0.082 \pm 0.009$ \\
SDSS-Calibration & 0.206 & 0.300 & 0.242 & 14.38 & 234 &$-0.26 \pm 0.03$ & $0.36 \pm 0.05$ & $0.153 \pm 0.014$ & $0.081 \pm 0.008$ \\
DES-Calibration & 0.206 & 0.300 & 0.243 & 14.24 & 351 & -0.26 & 0.38 $\pm$ 0.04 & 0.159 $\pm$ 0.004 & 0.067 $\pm$ 0.006 \\ \hline 
Bin 1  & 0.030 & 0.090 & 0.075 & 14.03 & 112 & -0.22 $\pm$ 0.05 & 0.52 $\pm$ 0.08 & 0.135 $\pm$ 0.020 & 0.070 $\pm$ 0.014 \\
Bin 2  & 0.090 & 0.130 & 0.112 & 14.02 & 203 & -0.25 $\pm$ 0.03 & 0.43 $\pm$ 0.06 & 0.168 $\pm$ 0.015 & 0.083 $\pm$ 0.009  \\
Bin 3  & 0.130 & 0.170 & 0.151 & 14.03 & 289 & -0.22 $\pm$ 0.03 & 0.45 $\pm$ 0.04 & 0.160 $\pm$ 0.012 & 0.075 $\pm$ 0.008  \\
Bin 4  & 0.170 & 0.210 & 0.187 & 14.17 & 260 & -0.25 $\pm$0.03 & 0.35 $\pm$ 0.05 & 0.159 $\pm$ 0.013 & 0.082 $\pm$ 0.008  \\
Bin 5 & 0.210 & 0.270 & 0.236 & 14.28 & 404 & -0.19 $\pm$0.02 & 0.36 $\pm$ 0.04 & 0.131 $\pm$ 0.011 & 0.079 $\pm$ 0.006  \\
Bin 6  & 0.270 & 0.360 & 0.307 & 14.24 & 385 & -0.18 $\pm$0.02 & 0.34 $\pm$ 0.04 & 0.120 $\pm$ 0.011 & 0.089 $\pm$ 0.006  \\
Bin 7  & 0.360 & 0.470 & 0.407 & 14.35 & 317 & -0.11 $\pm$0.03 & 0.33 $\pm$ 0.05 & 0.101 $\pm$ 0.011 & 0.053 $\pm$ 0.006  \\
Bin 8  & 0.470 & 0.600 & 0.528 & 14.35 & 353 & -0.08 $\pm$0.03& 0.33 $\pm$ 0.06 & 0.097 $\pm$ 0.012 & 0.092 $\pm$ 0.006  \\ \hline
TNG300-1 & 0.08 & 0.08 & 0.08 & 13.93 & 241 & $-0.37 \pm 0.04$ & $0.44 \pm 0.03$ & $0.132 \pm 0.011$ & $0.108 \pm 0.005$  \\
TNG300-1 no $\gamma$ & 0.08 & 0.08 & 0.08 & 13.93 & 241 & $0.13 \pm 0.02$ & $0.51 \pm 0.04$ &  & $0.138 \pm 0.006$  \\
TNG300-1 & 0.11 & 0.11 & 0.11 & 13.92 & 238 & $-0.39 \pm 0.05$ & $0.43 \pm 0.03$ & $0.138 \pm 0.012$ & $0.100 \pm 0.005$  \\
TNG300-1 no $\gamma$ & 0.11 & 0.11 & 0.11 & 13.92 & 238 & $0.12 \pm 0.02$ & $0.47 \pm 0.04$ &  & $0.126 \pm 0.006$  \\
TNG300-1 & 0.15 & 0.15 & 0.15 & 13.90 & 236 & $-0.34 \pm 0.04$ & $0.45 \pm 0.03$ & $0.127 \pm 0.010$ & $0.102 \pm 0.005$  \\
TNG300-1 no $\gamma$& 0.15 & 0.15 & 0.15 & 13.90 & 236 & $0.12 \pm 0.02$ & $0.47 \pm 0.04$ &  & $0.130 \pm 0.005$  \\
TNG300-1 & 0.18 & 0.18 & 0.18 & 13.91 & 235 & $-0.32 \pm 0.04$ & $0.44 \pm 0.03$ & $0.120 \pm 0.011$ & $0.102 \pm 0.005$  \\
TNG300-1 no $\gamma$ & 0.18 & 0.18 & 0.18 & 13.91 & 235 & $0.12 \pm 0.02$ & $0.47 \pm 0.04$ &  & $0.126 \pm 0.006$  \\
TNG300-1 & 0.24 & 0.24 & 0.24 & 13.87 & 234 & $-0.37 \pm 0.04$ & $0.46 \pm 0.03$ & $0.141 \pm 0.010$ & $0.094 \pm 0.004$  \\
TNG300-1 no $\gamma$ & 0.24 & 0.24 & 0.24 & 13.87 & 234 & $0.12 \pm 0.02$ & $0.46 \pm 0.04$ &  & $0.125 \pm 0.006$  \\
TNG300-1 & 0.31 & 0.31 & 0.31 & 13.84 & 233 & $-0.34 \pm 0.04$ & $0.44 \pm 0.03$ & $0.135 \pm 0.011$ & $0.106 \pm 0.005$  \\
TNG300-1 no $\gamma$ & 0.31 & 0.31 & 0.31 & 13.84 & 233 & $0.12 \pm 0.02$ & $0.45 \pm 0.04$ &  & $0.137 \pm 0.006$  \\
TNG300-1 & 0.40 & 0.40 & 0.40 & 13.80 & 231 & $-0.29 \pm 0.04$ & $0.48 \pm 0.03$ & $0.130 \pm 0.011$ & $0.108 \pm 0.005$  \\
TNG300-1 no $\gamma$ & 0.40 & 0.40 & 0.40 & 13.80 & 231 & $0.15 \pm 0.03$ & $0.47 \pm 0.04$ &  & $0.137 \pm 0.006$  \\
TNG300-1 & 0.52 & 0.52 & 0.52 & 13.71 & 236 & $-0.25 \pm 0.04$ & $0.49 \pm 0.03$ & $0.127 \pm 0.009$ & $0.111 \pm 0.005$  \\
TNG300-1 no $\gamma$& 0.52 & 0.52 & 0.52 & 13.71 & 236 & $0.18 \pm 0.03$ & $0.50 \pm 0.05$ &  & $0.149 \pm 0.007$  \\
\enddata
\label{tab:DESsim_comp}
\end{deluxetable*}
\end{appendix}

\end{document}